\documentclass[12pt]{article}
\usepackage{latexsym}
\usepackage{graphicx}
\usepackage{caption2}
\usepackage{psfrag}
\usepackage{amsmath}
\newcommand{\be}{\begin{equation}}
\newcommand{\ee}{\end{equation}}
\newcommand{\bea}{\begin{eqnarray}}
\newcommand{\eea}{\end{eqnarray}}

\newcommand{\nn}{\nonumber}
\newcommand{\lesssim}{ {\
\lower-1.2pt\vbox{\hbox{\rlap{$<$}\lower5pt\vbox{\hbox{$\sim$}}}}\ } }
\newcommand{\gtrsim}{ {\
\lower-1.2pt\vbox{\hbox{\rlap{$>$}\lower5pt\vbox{\hbox{$\sim$}}}}\ } }

\input epsf

\begin{document}

\begin{titlepage}

\begin{flushright}
\end{flushright}
\vspace*{1.5cm}
\begin{center}
{\Large \bf Duality violations and spectral sum rules}\\[2.0cm]

{\bf O. Cat\`{a}$^{a}$, M. Golterman$^{b}$ } and {\bf S. Peris$^{a}$}\\[1cm]

 $^{a}$Grup de F{\'\i}sica Te{\`o}rica and IFAE\\ Universitat
Aut{\`o}noma de Barcelona, 08193 Barcelona, Spain.\\[.5 cm]
 $^{b}$Department of Physics and Astronomy, San Francisco State
University\\1600 Holloway Ave, San Francisco, CA 94132, USA

\end{center}

\vspace*{1.0cm}

\begin{abstract}

We study the issue of duality violations in the $VV-AA$ vacuum polarization function
in the chiral limit.  This is done with the help of a  model with an expansion in
inverse powers of the number of colors, $N_c$, allowing us to consider resonances with
a finite width.  Due to these duality violations, the Operator Product Expansion
(OPE) and the moments of the spectral function (e.g. the Weinberg sum rules) do not
match at finite momentum, and we analyze this difference in detail. We also perform
a comparative study of many of the different methods proposed in the literature for
the extraction of the OPE parameters and find that, when applied to our model, they
all fare quite similarly. In fact, the model strongly suggests that a significant
improvement in precision can only be expected after duality violations are included.
To this end, we propose a method to parameterize these duality violations. The
method works quite well for the model, and  we hope that it may also be useful in
future determinations of OPE parameters in QCD.

\end{abstract}

\end{titlepage}

\section{Introduction}

The Operator Product Expansion (OPE) \cite{SVZ} is of paramount importance in QCD
because it is the short-distance window not only to fundamental parameters of the
Lagrangian such as quark masses and the coupling constant $\alpha_s$, but also to
other important quantities  not explicitly appearing in the Lagrangian, such as the
condensates. Moreover, there are cases, like that of the $\Pi_{VV}-\Pi_{AA}$ vacuum
polarization, where these condensates turn out to be related to
electroweak coupling constants governing the physics of kaon decay and its
associated CP violation \cite{relationship}. It is therefore essential to understand
how the OPE works and what its properties of convergence are. It is clear that using
the expansion in a region without  good convergence (or with no convergence at all)
may result in errors in the determination of these important parameters.

\begin{figure}
\renewcommand{\captionfont}{\small \it}
\renewcommand{\captionlabelfont}{\small \it}
\centering \psfrag{A}{$\mathrm{Re}(q^2)$}
\includegraphics[width=2.5in]{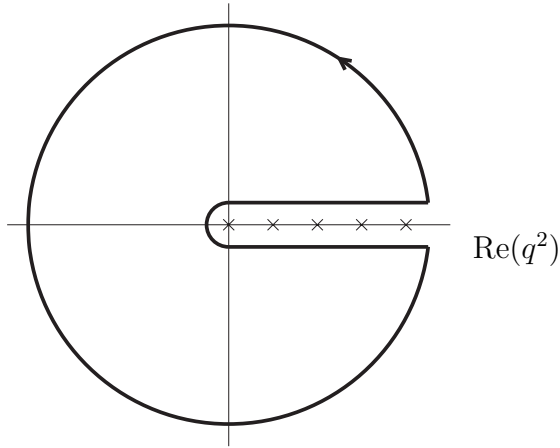}
\caption{This figure shows the contour in the complex plane leading to Eq.~(\ref{zero}).
The crosses on the real axis denote the position of the resonance
poles in the $N_c\rightarrow \infty$ limit, or the cut arising at finite $N_c$.
}\label{XXX}
\end{figure}

The OPE in QCD is believed to be only an asymptotic expansion in inverse powers of
$q^2$ in the complex $q^2$ plane  excluding the Minkowski region, ${\rm Re}~q^2>0$.
However, it is precisely in the Minkowski region where experimental data are
available, so that in order for the OPE to make contact with the experimental data,
one has to learn how to make the necessary rotation in the complex plane. A famous
trick \cite{Shankar}, relying on Cauchy's theorem, relates the integral of the
spectral function (i.e. of the experimental data) $\rho(t)=(1/\pi)~{\rm Im}\ \Pi(t)$
to that of the corresponding Green's function $\Pi(q^2)$ of complex argument $q^2$,
over the contour of Fig.~\ref{XXX}, chosen counter-clockwise. One
finds \cite{deRafael}
\begin{equation}\label{zero}
    \int_{0}^{s_0}dt\ P\left(t\right) \ \frac{1}{\pi}\ \mathrm{Im}\ \Pi(t)\ =
    -\frac{1}{2\pi i}\oint_{|q^2|=s_0}dq^2 \ P\left(q^2\right)\ \Pi(q^2),
\end{equation}
where $P(x)$ is any polynomial. As it stands, Eq.~(\ref{zero}) is of course an exact
mathematical statement. The right-hand side of Eq.~(\ref{zero}), however, contains
the full Green's function $\Pi(q^2)$ which is not available in QCD. It is then
assumed that, if the contour radius $s_0$ is large enough, use of the expansion
$\Pi_{OPE}(q^2)$ instead of the exact function will be a good enough approximation.
This is the rationale underlying all modern analyses in QCD based on the OPE such as
e.g. the interesting work of Ref.~\cite{Braaten}.

The approximation which consists of substituting $\Pi(q^2) \rightarrow
\Pi_{OPE}(q^2)$, with $s_0$ large enough, has been given the name \emph{duality}.
Duality refers to the fact that, if the approximation were exact and the
substitution $\Pi(q^2)\rightarrow \Pi_{OPE}(q^2)$ carried no error, one could say
that the integral of the experimental spectral function $\mathrm{Im}~\Pi(t)$ in
Eq.~(\ref{zero}) is dual to the OPE. The term \emph{duality violation},
consequently, refers to any contribution missed by this substitution.

The question is whether this approximation works for large $s_0$, and if so, how large
$s_0$ should be for duality to apply within a precision of, let us say, $10 \%$? In
order to address this question, it is essential to be able to assess the validity of
the approximation used, lest the extraction of OPE parameters not be afflicted by
uncontrolled systematic errors. Therefore, studying the contribution missed by the
replacement $\Pi(q^2)\rightarrow \Pi_{OPE}(q^2)$ in Eq.~(\ref{zero}) is a necessary
task in the present era of precision determination of QCD parameters.

That the OPE can at best only be an asymptotic expansion in the Euclidean region can
be most neatly seen in the case of a two-point Green's function made of a pair of
quark bilinears, such as  the vacuum polarization, in the large-$N_c$
limit~\cite{largeN}. This is because in this limit all meson resonances appear as
intermediate physical states with a vanishing width and, consequently, as poles on
the positive real axis in the $q^2$ complex plane. In contrast, the OPE is a series
expansion in inverse powers of $q^2$ (up to logarithms) with no sign of a pole
anywhere but at the origin, which means that it cannot converge on the positive real
axis. The OPE can thus at best be asymptotic,\footnote{A convergent expansion always
has a circular region of convergence characterized by a non-zero radius.} and even
then only provided the positive real axis, i.e. the Minkowski region, is excluded.
The region of validity of the expansion, therefore, is expected to be some sort of
angular sector around the origin not including the positive real axis. Much less
clear are more detailed properties such as, for instance, what the shape of this
angular sector may be.

Presently, the theory of duality violations in QCD is almost non-existing
and it is important to develop it. In the absence of such a theory, we see
no other choice but to resort to models.  Somewhat
surprisingly, not much work has been devoted to this issue. To our
knowledge, Ref.~\cite{Shifman} was the first to point out the
importance of duality violations. That the OPE may also miss other
important contributions has recently been suggested in Ref.~\cite{Zakharov}.

In this work we will study a \emph{model} of duality violations. We hasten to
emphasize that our model is \emph{not} QCD. In spite of this, we believe that it
shares enough properties with QCD to be an interesting theoretical laboratory for
studying issues related to duality violations. After all, our goal here is not a
detailed numerical study of duality violations in QCD,  but the much more modest one
of understanding what the main features are, and how these violations may manifest
themselves. This is not just an exercise of purely academic interest, however,
because the lessons learned from the model raise questions which \emph{must} also be
addressed in the case of QCD.   In addition, one of the results of our analysis will
be a possible method for unravelling duality violations in  QCD itself.

The model starts from  the  $\Pi_{VV}-\Pi_{AA}$ vacuum polarization Green's
function, in the chiral limit, given in terms of a spectrum consisting of two
infinite towers of equally-spaced meson resonances, together with the rho meson and
the massless pion, which are treated separately. Similar models can be found, e.g.,
 in Refs.~\cite{Shifman,GP01,GP02,Philyandco,espriu}.

The fact that the spectrum is infinite is crucial and underlies all the interesting
properties to be discussed. This infiniteness  of the spectrum is deeply rooted in
the $1/N_c$ expansion of QCD. In fact, as a first step, we will take the
large-$N_c$ limit and consider a spectrum of resonances with vanishing width. This
has the advantage that all quantities can be analytically calculated. As a second
step, following Ref.~\cite{Shifman}, we will give the resonances a width
as a subleading $1/N_c$ effect, while making sure that known analytical properties of
the Green's function are not spoiled. We will thus be able to study a rather
realistic model, allowing us to draw some conclusions on the issue
of duality violations in the real world.

It is obvious that, in order for Eq.~(\ref{zero}) to be useful, an assumption has to
be made. If the asymptotic regime starts at a value of $q^2$ which is higher than
the $s_0$ employed in this equation, the use of the approximate expansion
$\Pi_{OPE}(q^2)$ will completely miss all the physics of the full function
$\Pi(q^2)$, resulting in so large an error that, in fact, the approximation will
become totally invalid \cite{Donoghue}. In the case of our model
the asymptotic regime can only start after the
contribution of the two towers of resonances has been taken into account, which
means that the $s_0$ scale has to be larger than the two lowest-lying states in
these towers. The model might overemphasize the effects of duality violations,
but, by the same
token, it provides a means to identify the causes of these effects.

Past experience accumulated over the years suggests that the case of real QCD may be
less drastic than our model as far as the energy scale at which the OPE
starts being a useful approximation is concerned. In other words, previous analyses of QCD which
rely on the use of the OPE do not necessarily have to have systematic errors of the
same size as those encountered in our model. But the model can be used to compare
the different methods of analysis to see if there is one which is more reliable than
the others.

In real life, the values employed for $s_0$ are in the interval $\sim 1.5-3.0 \
\mathrm{GeV}^2$ and the lore is that this is expected to be high enough for the OPE
to set in. The agreement within present errors in the determination of $\alpha_s$
from tau decays \cite{Braaten} with those from other determinations at much larger
energy scales \cite{alpha} can be taken as a confirmation of this. However, a much
more accurate extraction in a new generation of precision experiments might show
discrepancies between the different methods of analysis.  In fact, there is already
a glimpse of trouble in the determination of quark masses \cite{masses}, which
becomes a more serious discrepancy in the terms which appear at higher order in the
OPE, such as condensates \cite{condensates}.\footnote{For a nice summary of results,
see Table 1 in Ref.~\cite{Narison}. Since certain electroweak matrix elements  are
related to condensates \cite{relationship}, one can also infer the latter from
lattice determinations of the former. See Table 8 in Ref.~\cite{lattice}.} We expect
the lessons on duality violations we learn from our model to be of help in all these
circumstances.

\section{Duality violations in the large-$N_c$ limit}

We will concentrate on the two-point  functions of vector and axial-vector currents
\be \label{correlator} \Pi^{V,A}_{\mu\nu}(q)=\ i\,\int d^4x\,e^{iqx}\langle
J^{V,A}_{\mu}(x) J^{\dag\ V,A}_{\nu}(0)\rangle = \left(q_{\mu} q_{\nu} - g_{\mu\nu}
q^2 \right)\Pi_{V,A}(q^2) \ , \ee
 where, for definiteness, $J_{V}^\mu(x) = {\overline d}(x)\gamma^\mu u(x)$ and $J_A^\mu(x) =
{\overline d}(x)\gamma^\mu \gamma^5 u(x)$. Throughout the paper, we will stay in the
chiral limit for simplicity. Both functions, $\Pi_{V,A}(q^2)$ in
Eq.~(\ref{correlator}) satisfy the dispersion relation \be \label{dispersion}
\Pi_{V,A}(q^2)= \int_0^{\infty} \frac{dt}{t-q^2-i\epsilon}\ \frac{1}{\pi}\ {\rm
Im}\,\Pi_{V,A}(t)\ , \ee
up to one subtraction.

According to the discussion in the introduction, and following
Refs.~\cite{Shifman,GP01}, we will assume the following spectra: \bea
\label{spectrum} \frac{1}{\pi}\  {\rm Im}\,\Pi_V(t)&=& 2 F_{\rho}^2
\delta(t-M_{\rho}^2) + 2 \sum_{n=0}^{\infty} F^2_V(n)\delta(t-M^2_V(n))\ ,\nonumber\\
\frac{1}{\pi}\  {\rm Im}\,\Pi_A(t)&=& 2 F_{0}^2 \delta(t) + 2 \sum_{n=0}^{\infty}
F^2_A(n)\delta(t-M^2_A(n))\ . \eea Here $F_\rho$ is the electromagnetic decay
constant of the $\rho$, and $M_\rho$ is its mass.  $F_{V,A}(n)$ are the
electromagnetic decay constants of resonance $n$ in the vector and axial channels,
while $M_{V,A}(n)$ are the corresponding masses. $F_0$ is the pion decay constant in
the chiral limit. The dependence on the resonance index $n$ is taken as follows:
\begin{equation}\label{two}
    F^2_{V,A}(n)=F^2= \mathrm{constant}\  ,
    \quad\quad M_{V,A}^2(n) = m_{V,A}^2 + n \ \Lambda^2\ ,
\end{equation}
which is compatible with known properties of the large-$N_c$ limit of QCD
\cite{largeN}, as well as properties of the associated Regge theory \cite{Regge}.

The combination
\begin{equation}\label{combo}
    \Pi(q^2)=\frac{1}{2}(\Pi_V(q^2)-\Pi_A(q^2))
\end{equation}
thus reads
\begin{equation}\label{one}
    \Pi(q^2)=\frac{F^2_0}{q^2}+\frac{F_{\rho}^2}{-q^2+M_{\rho}^2}-
    \sum_{n=0}^{\infty} \frac{F^2}{-q^2+M^2_A(n)}+ \sum_{n=0}^{\infty}
    \frac{F^2}{-q^2+M^2_V(n)}\ .
\end{equation}

Note that the upper end of the integration region in Eq.~(\ref{dispersion})
requires the introduction of a cutoff, $\Lambda_{\mathrm{CO}}$, to render the
integrals well defined. This cutoff must satisfy chiral symmetry in order
not to introduce any explicit breaking through the regulator. This is accomplished
by demanding that $\Lambda_{\mathrm{CO}}^2\sim N_V \Lambda^2 \sim N_A \Lambda^2 $ as
$N_{V,A}\rightarrow \infty$. Of course, physical results are independent of the
precise details as to how this limit is taken, in agreement with very general
properties of field theory \cite{GP02}.

One can see this explicitly in Eq.~(\ref{one}). The infinite sums are by itself
divergent and have to be regularized. In order to obtain a convergent result, one
has
  to choose the same regulator for both axial and vector towers, {\it{i.e.}}
  $N_V=N_A$, in this case. Infinities then cancel and the two-point
  function can be expressed in terms of the Digamma function  with the use of the identity
\begin{equation}\label{sum}
\lim_{N\rightarrow
\infty}\left\{\sum_{n=1}^{N}\frac{1}{z+n}-\sum_{n=1}^{N}\frac{1}{n}\right\}=-\psi(z)-\frac{1}{z}-
\gamma_E\ ,
\end{equation}
where $\gamma_E$ is the Euler-Mascheroni constant and $\psi(z)$ is the Digamma
function, defined as
\begin{equation}\label{definition}
\psi(z)=\int_0^{\infty} \left(\frac{e^{-t}}{t}-\frac{e^{-zt}}{1-e^{-t}} \right)\,
dt=\frac{d}{dz}\log{\Gamma(z)}\ .
\end{equation}
Thus, it follows that
\begin{equation}\label{onecompact}
    \Pi(q^2)=\frac{F^2_{0}}{q^2}+\frac{F_{\rho}^2}{-q^2+M_{\rho}^2}+
    \frac{F^2}{\Lambda^2}\left[\psi\left(\frac{-q^2+m_A^2}{\Lambda^2}\right)-
    \psi\left(\frac{-q^2+m_V^2}{\Lambda^2}\right)\right]\ .
\end{equation}
Furthermore, we will impose the
following conditions on our model:
\be \label{partonmodel} \frac{2}{3} \frac{N_c}{16\pi^2}=
\frac{F^2}{\Lambda^2}\ ,
\ee
so that the coefficient of the parton-model logarithm
is given by a free-quark loop;
\be \label{nodim2} F_{\rho}^2= F^2
\left(\frac{m_V^2}{\Lambda^2}-\frac{1}{2}\right) \quad , \quad  F_{0}^2 = F^2
\left(\frac{m_A^2}{\Lambda^2}-\frac{1}{2}\right)\ ,
\ee and
\be
\label{gluoncondensate}   -2 F_{\rho}^2 M_{\rho}^2 + F^2 \Lambda^2
    \left( \frac{m_V^4}{\Lambda^4} - \frac{m_V^2}{\Lambda^2}+\frac{1}{6}
\right) = F^2 \Lambda^2
    \left( \frac{m_A^4}{\Lambda^4} - \frac{m_A^2}{\Lambda^2}+\frac{1}{6}
\right) \ , \ee so that, as in the case of real QCD, there are effectively no
dimension-two or dimension-four operators in the OPE of $\Pi(q^2)$ (see below). Note
that Eqs.~(\ref{nodim2}) and (\ref{gluoncondensate}) would not be consistent if
$F_{\rho}=0$. Since we want our model to be semi-realistic, this is the reason why
we separated the rho meson from the vector tower in Eq.~(\ref{spectrum}). This is
also natural from the point of view of Regge theory, as the rho-meson mass is low
enough to be below the onset of Regge trajectories (presumably at a scale of
$\mathcal{O}(1\ GeV)$). The $a_1$ is heavier and thus we include it in the axial
tower.\footnote{We could also have separated the $a_1$, but that would introduce
unnecessary free parameters without any crucial change in the analytic properties of
the OPE.} At any rate, we will see that what really matters is the analytic
properties of the OPE and these depend on the tower as a whole, rather than on the
properties of any individual resonance.

Knowledge of the spectrum (\ref{spectrum}) allows us to calculate the full Green's
function $\Pi(q^2)$ explicitly. Defining the spectral function
$\rho(t)=(1/\pi)~\mathrm{Im}~\Pi(t)$ one has that (with $Q^2=-q^2$)
\begin{equation}\label{three}
    \Pi(-Q^2)=\int_0^{\infty} \frac{dt}{t+Q^2-i\epsilon}\ \rho(t)\qquad
    \Longrightarrow_{_{\!\!\!\!\!\!\!\!\!\!\!\!\!\!\!\!\!\!\!\!\!\mathrm{large}\ Q^2>0}}
 \qquad  \Pi_{OPE}(-Q^2) \approx \sum_{k=1,2,3,...}
   \frac{C_{2k}}{Q^{2k}}\ ,
\end{equation}
where the $1/Q^2$ expansion in the second equation above is akin to the Operator
Product Expansion in QCD. Note that a naive expansion in powers of $t/Q^2$ under the
integral sign is \emph{not} an allowed mathematical operation. Consequently, in
order to obtain the correct $1/Q^2$ expansion due care must be exercised.  The correct
OPE can be derived by using (the derivative of) Eq.~(\ref{definition}) and the
expansion
\be e^{xz}\frac{z}{e^z -1}=\sum_{n=0}^{\infty}B_{n}(x)\frac{1}{n!}z^n\ ,
\ee
wherein $B_n(x)$ are the Bernoulli polynomials. In this way one arrives at the
following expressions for the OPE coefficients $C_{2k}$:
\begin{eqnarray}
\label{ope}
     C_{2k} &=& -F_{0}^2 \ \delta_{k,1}+\nonumber \\
     &&\!\!\!\!\!\!\!\!\!\!\!\!
     (-1)^{k+1} \left[ F_{\rho}^2M_{\rho}^{2k-2}  -\frac{1}{k}F^2 \Lambda^{2k-2}
  \left\{B_{k}\left(\frac{m_V^2}{\Lambda^2}\right) -
  B_{k}\left(\frac{m_A^2}{\Lambda^2}\right)\right\}\right]\ .
\end{eqnarray}
In particular, the first few coefficients read\footnote{In fact, $C_{2,4}$ vanish
identically because of the conditions (\ref{partonmodel},\ref{gluoncondensate}).}
\begin{eqnarray}\label{examples}
  C_2 &=& + F_{\rho}^2 - \,\, F_{0}^2 -\,\,
  F^2\left\{B_1\left(\frac{m_V^2}{\Lambda^2}\right) -
  B_1\left(\frac{m_A^2}{\Lambda^2}\right)\right\}\nonumber \ ,\\
  C_4 &=&  - F_{\rho}^2 M_{\rho}^2+
  \frac{1}{2}F^2 \Lambda^2\left\{B_2\left(\frac{m_V^2}{\Lambda^2}\right) -
  B_2\left(\frac{m_A^2}{\Lambda^2}\right)\right\}\nonumber \ ,\\
  C_6 &=&  + F_{\rho}^2 M_{\rho}^4-
  \frac{1}{3}F^2 \Lambda^4\left\{B_3\left(\frac{m_V^2}{\Lambda^2}\right) -
  B_3\left(\frac{m_A^2}{\Lambda^2}\right)\right\} \nonumber \ ,\\
  C_8 &=&  - F_{\rho}^2 M_{\rho}^6+
  \frac{1}{4}F^2 \Lambda^6\left\{B_4\left(\frac{m_V^2}{\Lambda^2}\right) -
  B_4\left(\frac{m_A^2}{\Lambda^2}\right)\right\} \ ,
  \end{eqnarray}
and the first few Bernoulli polynomials are \bea B_{0}(x)  =  1 \quad , \quad
B_{1}(x) = x-\frac{1}{2}\quad , \quad
B_{2}(x)  =  x^2-x+\frac{1}{6}\,, \nn \\
B_{3}(x)  =  x^3 -\frac{3}{2}x^2+\frac{1}{2}x\quad , \quad B_{4}(x)  =
x^4-2x^3+x^2-\frac{1}{30}\ . \eea

As already stated, the OPE cannot be obtained by naively expanding at large $Q^2$
under the integral sign in Eq.~(\ref{three}). We emphasize that this feature is not
specific to our model and is, in fact, also true in QCD. Note that, if a naive
$1/Q^2$ expansion were valid in QCD, there could be no $\log Q^2$ dependence in
$C_{2k}$ and all anomalous dimensions would have to vanish, which is obviously not
the case. Our model does share with QCD the property of a non-trivial expansion in
$1/Q^2$ but, in the large-$N_c$ limit, it still misses all these log's. With the
introduction of finite widths in the next section, this drawback will be somewhat ameliorated.

It thus follows from the previous discussion that the moments
\begin{equation}\label{five}
    M_{n}(s_0)=\int_{0}^{s_0}\ dt \ t^{n} \rho(t)
\end{equation}
are \emph{not} the coefficients of the OPE, $C_{2k}$. In order to obtain
the actual relation between $M_{n}(s_0)$ and $C_{2k}$, it is convenient to
first calculate the Laplace transform of the dispersion relation (\ref{three})
\begin{equation}\label{six}
    \Pi(-Q^2)= \int_{0}^{\infty} d\tau\ e^{- \tau Q^2}\ \widehat{\rho}(\tau)\quad
    \mathrm{with}\quad
\widehat{\rho}(\tau)= \int_{0}^{\infty} dt\ e^{-t\tau}\ \rho(t)\ .
\end{equation}
Plugging in the $1/Q^2$ expansion of Eq.~(\ref{three}) and identifying Laplace
transforms on both sides, one obtains
\begin{equation}\label{eight}
    \sum_{k=1}^{\infty} \frac{C_{2k}}{(k-1)!}\ \tau^{k-1}=
    \int_{0}^{\infty} dt\ e^{-t\tau} \rho(t)\ ,
\end{equation}
and therefore, finally,
\begin{equation}\label{nine}
    C_{2k}= \lim_{\tau\rightarrow 0}\left\{(-1)^{k-1}\int_{0}^{\infty} dt\ t^{k-1}\ e^{-t\tau}
    \rho(t)\right\}\ .
\end{equation}
As an exercise one can use the expression (\ref{nine}) to recover our results in
Eq.~(\ref{examples}). It is the slow fall-off with $t$ of $\rho(t)$ which is
responsible for all the complications and, in particular, the duality violations. If
$\rho(t)$ were falling off faster than any power at large $t$, one could take the
$\tau\rightarrow 0$ inside the integral, and the naive $1/Q^2$ expansion
of the dispersion integral in Eq.~(\ref{three}) would be valid.

In actual fact, the calculation of the moments (\ref{five})
yields
\begin{eqnarray}\label{eleven}
  M_0(s_0) &=& C_2-F^2 \Big[B_1(x_V)-B_1(x_A)\Big]\nonumber \ , \\
  M_1(s_0) &=&- C_4 -F^2 \Big[B_1(x_V)-B_1(x_A)\Big]s_0 + \frac{1}{2}F^2\Lambda^2
  \Big[B_2(x_V)-B_2(x_A)\Big]\nonumber \ , \\
  M_2(s_0) &=& C_6 - F^2 \Big[B_1(x_V)-B_1(x_A)\Big]s_0^2+
  F^2\Lambda^2\Big[B_2(x_V)-B_2(x_A)\Big] s_0 \nonumber\\
  &&- \frac{1}{3} F^2 \Lambda^4 \Big[B_3(x_V)-B_3(x_A)\Big] \nonumber \ ,\\
  M_3(s_0) &=& -C_8 -F^2  \Big[B_1(x_V)-B_1(x_A)\Big] s_0^3+
  \frac{3}{2} F^2\Lambda^2  \Big[B_2(x_V)-B_2(x_A)\Big] s_0^2 \nonumber\\
&&- F^2\Lambda^4 \Big[B_3(x_V)-B_3(x_A)\Big] s_0 +\frac{1}{4}F^2 \Lambda^6
\Big[B_4(x_V)-B_4(x_A)\Big]\ ,
\end{eqnarray}
where the $C_{2k}$ are given in Eqs.~(\ref{examples}), and we
have defined $0 < x_{V,A}< 1$ as the fractional parts of $(s_0-m^2_{V,A})/\Lambda^2$,
respectively. The above expressions are valid
when $s_0>m_V^2$, i.e., when both the vector and axial towers are included.
In these expressions, the points $x_{V,A}=0,1$ should be excluded
because they correspond to the singularities of the delta functions in the spectrum,
cf. Eq.~(\ref{spectrum}). We have kept the coefficients $C_{2,4}$ for illustrative
purposes, even though they actually vanish because of the constraints
(\ref{partonmodel}-\ref{gluoncondensate}). Perhaps several comments to explain the
pattern in the result of Eq.~(\ref{eleven}) are in order.

\begin{figure}
\renewcommand{\captionfont}{\small \it}
\renewcommand{\captionlabelfont}{\small \it}
\centering
\includegraphics[width=2.5in]{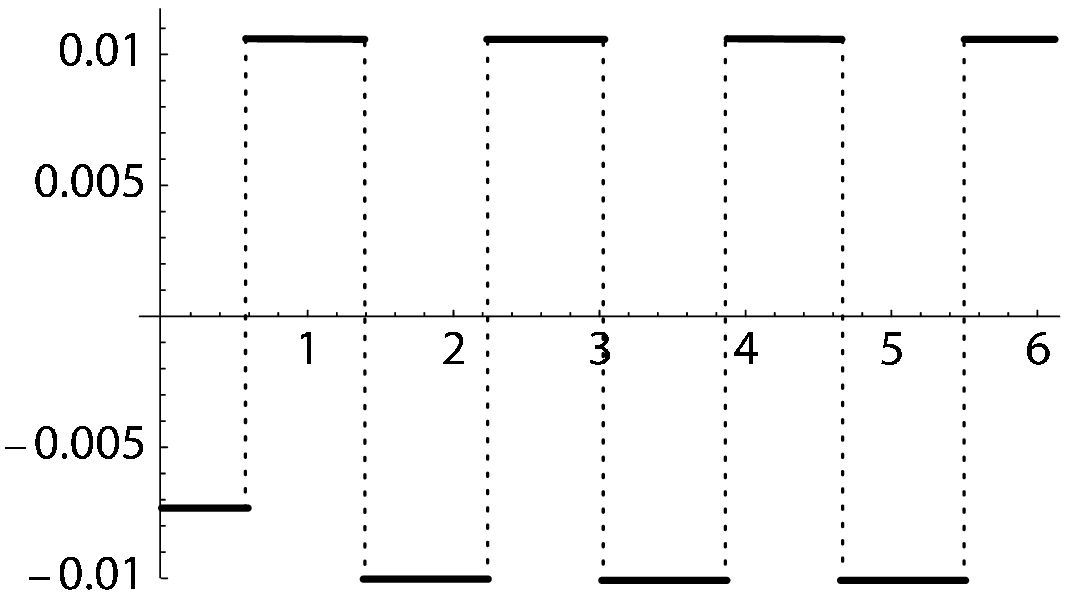}
\hspace {1cm}
\includegraphics[width=2.5in]{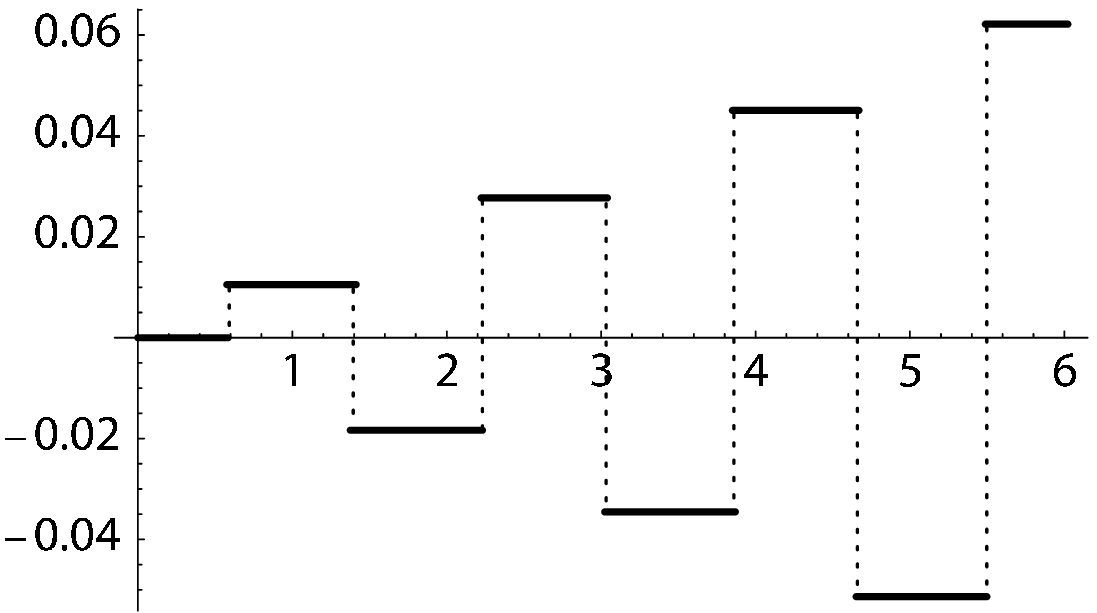}\\
\vspace {1cm}
\includegraphics[width=2.5in]{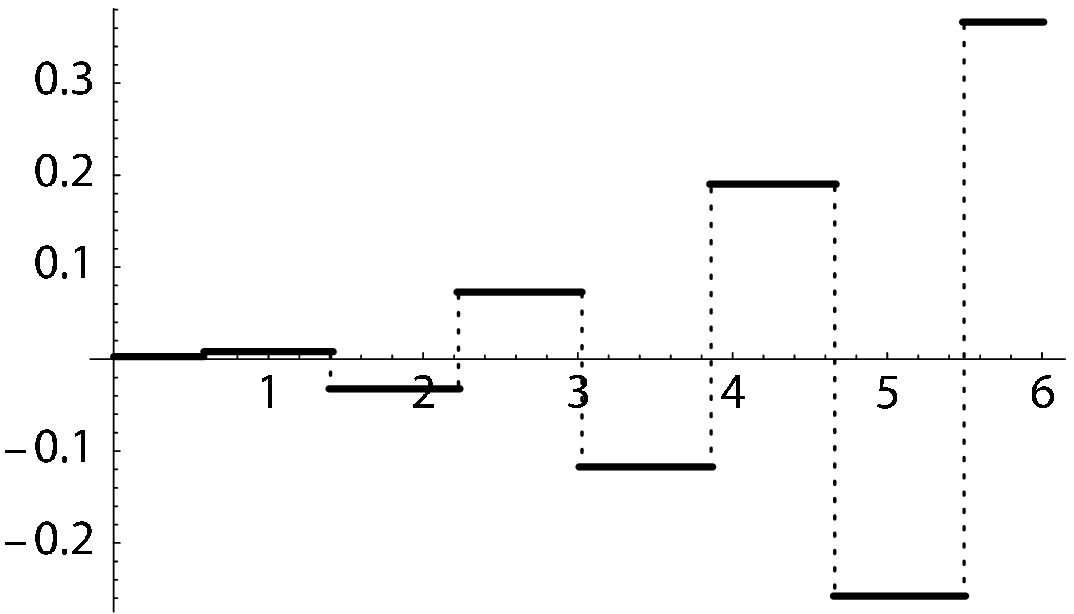}
\hspace {1cm}
\includegraphics[width=2.5in]{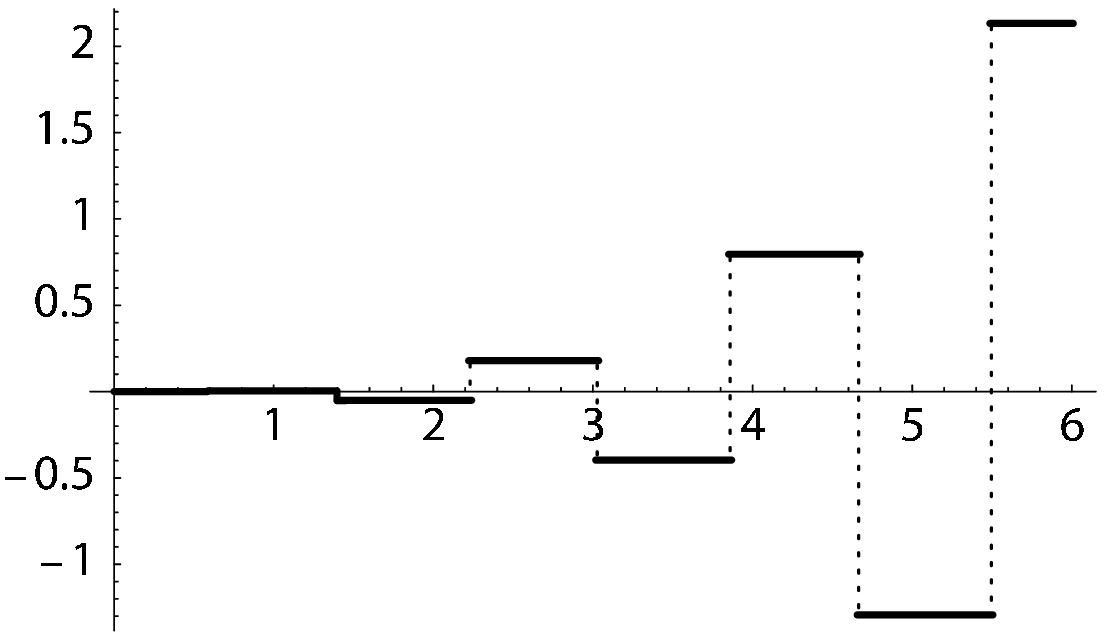}
\caption{From left to right: $M_0$, $M_1$ (first row), $M_2$ and $M_3$ (second row),
defined in Eq.~(\ref{five}), as functions of the energy scale, $s_0$ (in
$\mathrm{GeV^2}$).}\label{fig1}
\end{figure}

First, these expressions clearly show that there is no $s_0$ at which the moments
equal the OPE coefficients, $C_{2k}$, and consequently, in the world of
$N_c\rightarrow \infty$ described by our model, there is no such thing as a true
``duality point.'' There are always duality violations in the form of a polynomial
in the variables $x_{V,A}$. As can be seen in Fig.~\ref{fig1},  these violations are
step-function ``oscillations'' around the corresponding OPE coefficient. Note that
the OPE coefficients $C_{6,8}$ are numerically negligible (using the parameters in
Eq.~(\ref{nature}) below) compared to the oscillations shown in Fig.~\ref{fig1} for
the corresponding moments $M_{2,3}$.

Looking at Eqs.~(\ref{eleven}), one sees that there is some recursiveness, i.e.
higher moments depend on the duality violation already existing in lower moments.
Also, higher moments have higher powers of $s_0$ modulating their duality
violations, so the oscillations grow larger for higher moments. This means that the
larger the $s_0$ at which one is computing the moments (\ref{five}), the worse the
duality violation. All this behavior is clearly shown in Fig.~\ref{fig1}, where we
have taken as the values for our parameters the set\footnote{This  choice is
dictated by the fact that, when a nonvanishing width will be included in the next
section, the spectral function resembles qualitatively the experimental data. See
Fig.~\ref{fig3} below.  This choice of parameters also satisfies
Eqs.~(\ref{partonmodel}-\ref{gluoncondensate}).}
\begin{eqnarray}\label{nature}
\hspace{-2. cm} F_{0}= 85.8 \,\,{\mathrm{MeV}}\ , \quad F_{\rho}= 133.884
\,\,{\mathrm{MeV}}\ ,
\quad F= 143.758 \,\,{\mathrm{MeV}}\, ,\qquad \qquad \\
M_{\rho}= 0.767 \,\,{\mathrm{GeV}}, \quad m_A= 1.182 \,\,{\mathrm{GeV}}, \quad m_V=
1.49371 \,\,{\mathrm{GeV}}\, , \quad \Lambda= 1.2774 \,\, {\mathrm{GeV}}\ .\nonumber
\end{eqnarray}

Second, as we have said, the end points $x_{V,A}=0,1$ are to be excluded from the
result shown in Eq.~(\ref{eleven}). They correspond to the singularities of the
delta functions in Eq.~(\ref{spectrum}), i.e. to the position of the meson masses,
and cause the appearance of step-like discontinuities at the same location in all
the moments $M_n(s_0)$. At the jump produced by, e.g., the kick of a vector
resonance after the drop produced by an axial-vector one, Fig.~\ref{fig1}  shows how
one crosses the value of the corresponding coefficient of the OPE. However, since
the slope at these points is infinite, there is no way to predict the $C_{2k}$
coefficient from the value of the corresponding moment.

One might think that the introduction of a finite width could  drastically improve
the present situation at large $N_c$. However, this is not really true in the model.
Although introducing a finite width certainly changes the slope to make it finite,
it also moves the location of the duality points from one moment to the next. As we
will discuss in more detail in the next section, the net result is that it is still
very difficult to make accurate predictions.

The best way to try to understand the relationship between the moments and the
coefficients of the OPE is through the use of Cauchy's theorem, Eq.~(\ref{zero}). As
already discussed in the introduction, the replacement $\Pi(q^2)\rightarrow
\Pi_{OPE}(q^2)$ would be valid if the OPE were a convergent expansion with a radius
of convergence including the circumference $|q^2|=s_0$. However, in QCD the OPE is
not a convergent expansion anywhere on the ${\rm Re}~q^2>0$ axis. Since this crucial
property is shared by our model, we can now use the model to study how
Eq.~(\ref{zero}) should be modified when the expansion $\Pi_{OPE}(q^2)$ is used
instead of the exact function, $\Pi(q^2)$.

To this end, let us define $\mathcal{D}^{[n]}(s_0)$ from the equation
\begin{equation}\label{ddelta}
    \int_{0}^{s_0}dt\ t^{n}\ \rho(t)\ =
    -\frac{1}{2\pi i}\oint_{|q^2|=s_0}dq^2 \ q^{2n}\
    \Pi_{OPE}(q^2)+ \mathcal{D}^{[n]}(s_0)\ .
\end{equation}
Obviously, the term $\mathcal{D}^{[n]}(s_0)$  measures the amount of duality
violation and we would like to obtain a more explicit expression for it in our model.

In order to do this, note that the OPE in our model is related to the large $|z|$
expansion of the Digamma function,
\begin{equation}\label{asympt}
\psi(z)\sim \log{z}-\frac{1}{2z}-\sum_{n=1}^{\infty}\frac{B_{2n}}{2n\ z^{2n}}\quad
,\qquad |\arg(z)|<\pi \ ,
\end{equation}
which is not a valid expansion on the negative real axis in the complex $z$ plane.
We may, nevertheless, obtain a valid expansion in this region if we use the
``reflection" property of the Digamma function
\begin{equation}\label{reflexion}
    \psi(z)=\psi(-z)-\pi\ \cot(\pi z)-\frac{1}{z}\ ,
\end{equation}
which is valid for any complex number $z$. Using Eq.~(\ref{reflexion}), the
appropriate expansions for large values of $|q^2|$ in the whole $q^2$ plane then read
\begin{equation}\label{DVinf}
\Pi(q^2)\approx \Bigg\{
\begin{array}{ll}
    \!\!\!\Pi_{OPE}(q^2) +\mathcal{O}
    \left( e^{-2 \pi|q^2|/\Lambda^2}\right)& ,
    \  \hbox{${\rm Re}\ q^2\leq0$\ ;} \\
    \!\!\!\Pi_{OPE}(q^2)+\Delta_\infty(q^2)+
    \mathcal{O}\left( e^{-2 \pi|q^2|/\Lambda^2}\right)&,\  \hbox{${\rm Re}\ q^2\geq0$\ ,} \\
\end{array}
\end{equation}
where
\begin{equation}\label{delta}
    \Delta_\infty(q^2)=\frac{\pi F^2}{\Lambda^2}\ \Bigg\{\cot \left[\pi\frac{-q^2+m_V^2}{\Lambda^2}
    \right]- \cot \left[\pi\frac{-q^2+m_A^2}{\Lambda^2}\right]\!\Bigg\}\ ,
\end{equation}
and the subscript ``$\infty$'' refers to the $N_c\rightarrow \infty$ limit. Let us
emphasize again that the appearance of $\Delta_\infty(q^2)$ in Eq.~(\ref{DVinf})
gives rise to a violation of duality which does \emph{not} go away as
$|q^2|\rightarrow \infty$ (in particular it contains simple poles on the positive
real axis in the $q^2$ plane). This duality violation comes on top of the expected
exponentially suppressed contributions, which we have exhibited explicitly in
Eq.~(\ref{DVinf}), and which originate from the fact that the OPE, in those regions
where it is valid, is only an asymptotic expansion.\footnote{See the Appendix.}

Consequently, up to an exponentially small contribution, the duality violation
$\mathcal{D}$ term can be written as an integral over the semicircle $|q^2|=s_0$
with $\mathrm{Re}~q^2\geq 0$  (taken counter-clockwise) of the function
$\Delta_{\infty}(q^2)$,
\begin{equation}\label{dualviol}
    \mathcal{D}^{[n]}(s_0)=-\frac{1}{2\pi i}\
    \int_{_{_{_{\!\!\!\!\!\!\!\!\!\!\!\!\!\!\!\!\!\!{\begin{array}{c}
      |q^2|=s_0 \\ {\rm Re}\ q^2\geq 0
    \end{array}}}}}} \!\!\!\!\!\!\!\!\!\!\!\!\!\!\! dq^2\ q^{2n}\
    \Delta_\infty(q^2)
    + \mathcal{O}\left( e^{-2\pi s_0/\Lambda^2}\right)\ .
\end{equation}

\begin{figure}
\renewcommand{\captionfont}{\small \it}
\renewcommand{\captionlabelfont}{\small \it}
\centering \psfrag{A}{$\mathrm{Re}(q^2)$} \psfrag{B}{$+i\,s_0$}
\psfrag{C}{$-i\,s_0$} \psfrag{D}{$-i\,\epsilon$} \psfrag{F}{$+i\,\epsilon$}
\includegraphics[width=2.3in]{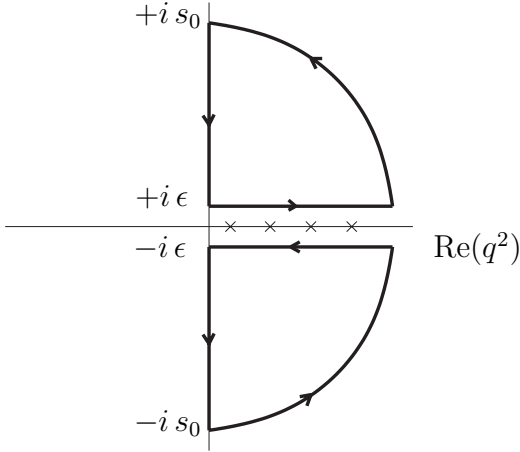}\caption{Contour used for the definition of the functions
$\mathcal{D}^{[n]}_{oscill.}(s_0)$ and $\mathcal{D}^{[n]}_{const.}(s_0)$ in Eqs.~(\ref{oscill},\ref{const}).}\label{YYY}
\end{figure}

Furthermore, Eq.~(\ref{dualviol}) can be rewritten in the form of a spectral
integral in the following way. The integral over the contour in Fig.~\ref{YYY}
vanishes, since it encloses no singularity. It follows that the duality violation
function $\mathcal{D}^{[n]}(s_0)$ in Eq.~(\ref{dualviol}) can also be expressed, up
to exponentially small terms $\mathcal{O}(e^{-2 \pi s_0/\Lambda^2})$, as
\begin{equation}\label{dualspectral}
    \mathcal{D}^{[n]}(s_0)=\mathcal{D}^{[n]}_{oscill.}(s_0)+\mathcal{D}^{[n]}_{const.}(s_0)\ ,
\end{equation}
with
\begin{eqnarray}
\mathcal{D}^{[n]}_{oscill.}(s_0)&=&\int_0^{s_0}  dt\ t^n
\ \frac{1}{\pi}\ \mathrm{Im} \ \Delta_\infty(t+i\varepsilon)\label{oscill}\\
   \mathcal{D}^{[n]}_{const.}(s_0)&=& - \frac{1}{2\pi i}\left\{\int_{-is_0}^{-i\epsilon}
   +\int_{i\epsilon}^{is_0}\right\}
   dq^2\ q^{2n}\ \Delta_\infty(q^2)\label{const}\ .
\end{eqnarray}
In these expressions, the integral (\ref{oscill}) is taken over the positive real
axis, in the complex $q^2$ plane, whereas the integrals (\ref{const}) are taken over
the imaginary axis (with $\epsilon\rightarrow 0^+$).  We see that
expression (\ref{oscill}) is of the form of a spectral integral, just like the left-hand
side of Eq.~(\ref{ddelta}). In fact, for $N_c\rightarrow \infty$,
$\mathrm{Im}~\Delta_\infty(q^2+i\varepsilon)$ is also a sum of delta
functions with poles precisely located according to the spectrum given by the towers
of Eqs.~(\ref{two}). This follows from the expansion for
the cotangent,
\begin{equation}\label{cot}
    \pi \cot\left(\pi z\right)=\frac{1}{z}+ 2z\sum_{n=1}^{\infty}\frac{1}{z^2-n^2}\ .
\end{equation}

Physically, the function $\mathcal{D}^{[n]}_{oscill.}(s_0)$ is responsible for
balancing all the oscillations which are present in the spectral integral on the
left-hand side of Eq.~(\ref{ddelta}) but which are missed by the contour integral of
the OPE on the right-hand side. The function $\mathcal{D}^{[n]}_{const.}(s_0)$, on
the other hand, does not have any oscillating component, and is dominated by
the region of small $q^2$, decaying exponentially fast in $s_0$  to a constant. One has
\begin{equation}\label{counterterms}
    \mathcal{D}^{[n]}_{const.}(s_0)\ = \ \mathcal{C}^{[n]}_\infty +
    \ \mathcal{O}\left(e^{-2 \pi s_0/\Lambda^2}\right)\ ,
\end{equation}
with,  in our $N_c=\infty$ model,\footnote{In order to obtain this
result, note that in our model $1<\frac{m_V^2}{\Lambda^2}<2$,
whereas $\frac{m_A^2}{\Lambda^2}<1$, cf. Eq.~(\ref{nature}).}
\begin{equation}\label{constants}
\mathcal{C}^{[n]}_\infty=\frac{F^2\Lambda^{2n}}{n+1}\left\{B_{n+1}\left(\frac{m_V^2}{\Lambda^2}\right)-
  B_{n+1}\left(\frac{m_A^2}{\Lambda^2}\right)\right\}
  -  F^2\left(m_V^2-\Lambda^2\right)^n \ .
\end{equation}
Using this result for $n=0,1,2,3$ in Eq.~(\ref{const}) together with
Eqs.~(\ref{oscill}), (\ref{dualspectral}) and (\ref{ddelta}) one may recover the results of
Eq.~(\ref{eleven}) for the moments $M_{0,1,2,3}$.  Again, the total
contribution to $\mathcal{D}^{[n]}(s_0)$ in Eq.~(\ref{ddelta}) does not vanish, even
in the $s_0\rightarrow \infty$ limit.

\begin{figure}
\renewcommand{\captionfont}{\small \it}
\renewcommand{\captionlabelfont}{\small \it}
\centering
\includegraphics[width=2.5in]{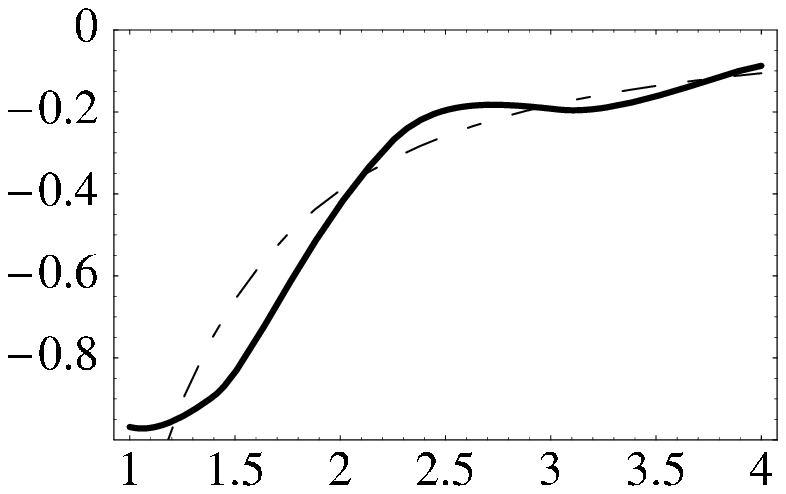}
\hspace {1cm}
\includegraphics[width=2.5in]{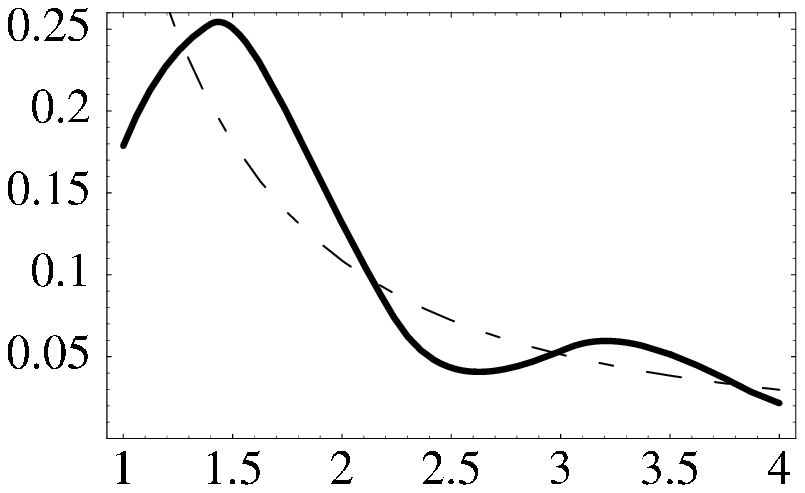}\\
\vspace {0.5cm}
\includegraphics[width=2.5in]{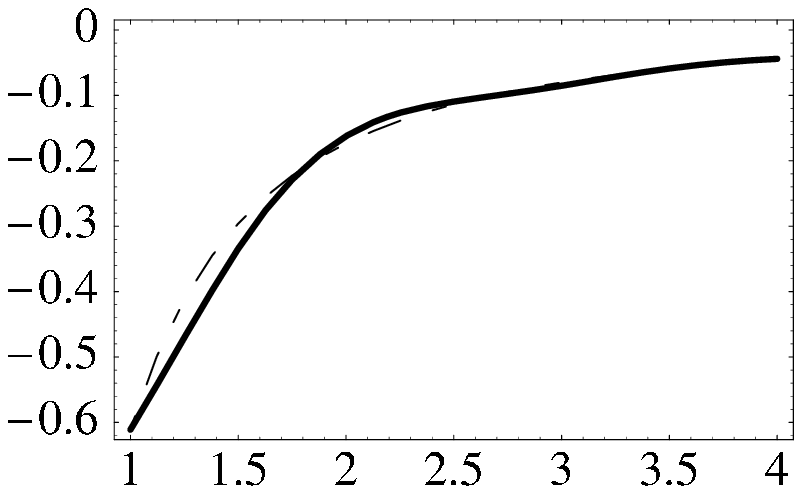}
\caption{Integrals of the pinched weights, $\int_0^{s_0}dt\ P_{1,2,3}(t)
\rho(t)$ defined in
the text (solid lines, from left to right and top to bottom) and the corresponding
OPE prediction (dashed lines) as a function of $s_0$ (in $\mathrm{GeV^2}$). These
integrals are expressed in units of $10^{-2}~{\rm GeV}^2$. }\label{ZZ}
\end{figure}

Since duality violations are caused by the lack of convergence near
the positive real axis in the $q^2$ plane, a possible  strategy to eliminate them
might be to use suitable polynomials $P(t)$ in Eq.~(\ref{ddelta}) which suppress
the contribution precisely from this region in $q^2$ by having a zero at $q^2=s_0$.
Such polynomials have been referred to as ``pinched weights'' in
Ref.~\cite{Cirigliano}.

This is confirmed by our large-$N_c$ model, and it is related to the recursiveness
present in the different moments in Eq.~(\ref{eleven}). For instance, a pinched
weight of the form  $(1-t/s_0)$ automatically cancels the duality violating
oscillation proportional to $B_1(x_V)-B_1(x_A)$ between the  moments $M_0$ and $M_1$.
The cancelation, however, is not complete and this pinched weight leaves behind an
oscillation proportional to $[B_2(x_V)-B_2(x_A)]/s_0$ which, although suppressed at
high $s_0$, does not vanish completely.

One can actually exploit this feature a little further and, by inspection of
Eq.~(\ref{eleven}), one easily sees that $P_3=(1-t/s_0)^3$ is the pinched weight
which best suppresses the duality violating oscillations among the moments
$M_{0,1,2,3}$ since it only leaves behind a term proportional to
$[B_4(x_V)-B_4(x_A)]/s_0^3$. This is illustrated in Fig.~\ref{ZZ}, where it is shown
how much of a reduction in the oscillations of the left-hand side of
Eq.~(\ref{ddelta}) is achieved by the pinched weight $P_3$ in comparison with
other pinched weights previously proposed in the literature \cite{Cirigliano}, i.e.
$P_1= (1-3 t/s_0) (1-t/s_0)^2$ and $P_2=(t/s_0) (1-t/s_0)^2$.

Figure~\ref{ZZ} also shows that the OPE curve more or less interpolates through the
oscillations in the spectral integral. As one can see, there are points at which the
oscillation happens to cross the OPE curve so that, at these points,  all  duality
violations effectively vanish. Based on this, one could imagine guessing the OPE
coefficients by some sort of averaging over one full oscillation. Regretfully, this
behavior sensitively depends on the fact that we are considering the $N_c\rightarrow
\infty$ limit. The inclusion of a finite width essentially wipes out any oscillating
signal and one is left without any clue as to where to take the average.

In the next section we will discuss in detail the inclusion of a finite width as
well as those features which survive the passage to this more realistic setting.

\section{Duality violations at finite $N_c$}

Since the issue of duality violations hinges strongly on the analytic properties of
the OPE and the corresponding Green's functions, it is clear that the introduction
of a finite width has to be done with some care in order not to spoil these analytic
properties.

Moreover, in QCD finite widths appear as a consequence of subleading effects in the
$1/N_c$ expansion. It is clear that if one knew how to properly include all these
$1/N_c$ effects, one would incorporate just the right $1/N_c$ corrections to give
resonances a finite width, while at the same time preserving the relevant analytic
properties of $\Pi(q^2)$. Among those properties, the most important one for us will
be the existence of a cut in the complex plane only for $\mathrm{Re}~q^2>0$. A
simple Breit-Wigner function, for instance, is not good enough.

However, the solution of QCD in the large-$N_c$ limit is unknown, let alone the
$1/N_c$ corrections, so it is clear that one has to model these effects in some
simple way in order to make progress. This is what has been done in
Ref.~\cite{Shifman}. The hope is that the conclusions drawn from such a model will not
depend in any crucial way on the particular details but rather on the generic
properties of the model.

With this philosophy in mind we follow Ref.~\cite{Shifman} and introduce a
subleading $1/N_c$ width, writing the vacuum polarization  $\Pi(q^2)$ as
\begin{equation}\label{fifteen}
    \Pi(q^2)=\left(1- \frac{a}{\pi N_c}\right)^{-1}
    \left\{- \frac{F^2_{0}}{z}+\frac{F_{\rho}^2}{z+M_{\rho}^2}- \sum_{n=0}^{\infty}
    \frac{F^2}{z+M_A^2(n)}+\sum_{n=0}^{\infty} \frac{F^2}{z+M_V^2(n)}\right\}\ ,
\end{equation}
where now the variable $z$ denotes the combination
\begin{equation}\label{sixteen}
    z=\Lambda^2 \left(\frac{-q^2-i\epsilon}{\Lambda^2}\right)^{\zeta}\quad , \quad
    \zeta=1-\frac{a}{\pi N_c}\quad ,
\end{equation}
and $M_{V,A}(n)$ are given in Eq.~(\ref{two}). As in the zero-width case, the
infinite sums can be expressed in terms of the Digamma function, and one obtains
\begin{equation}\label{fifteencompact}
\Pi(q^2)=\frac{1}{\zeta}
    \left\{-\frac{F^2_{0}}{z}+\frac{F_{\rho}^2}{z+M_{\rho}^2}+
    \frac{F^2}{\Lambda^2}\left[\psi\left(\frac{z+m_A^2}{\Lambda^2}\right)-
    \psi\left(\frac{z+m_V^2}{\Lambda^2}\right)\right]\right\}\ .
\end{equation}
Obviously, in the $N_c\rightarrow\infty$ limit we recover the results of the
previous section. The conditions of Eqs.~(\ref{partonmodel}), (\ref{nodim2}) and
(\ref{gluoncondensate}) impose now the absence of any $\log z, 1/z$ and $1/z^2$
terms in the large $Q^2=-q^2>0$ expansion of the Green's function $\Pi(q^2)$ in
Eq.~(\ref{fifteencompact}).

\begin{figure}
\renewcommand{\captionfont}{\small \it}
\renewcommand{\captionlabelfont}{\small \it}
\centering
\includegraphics[width=2.5in]{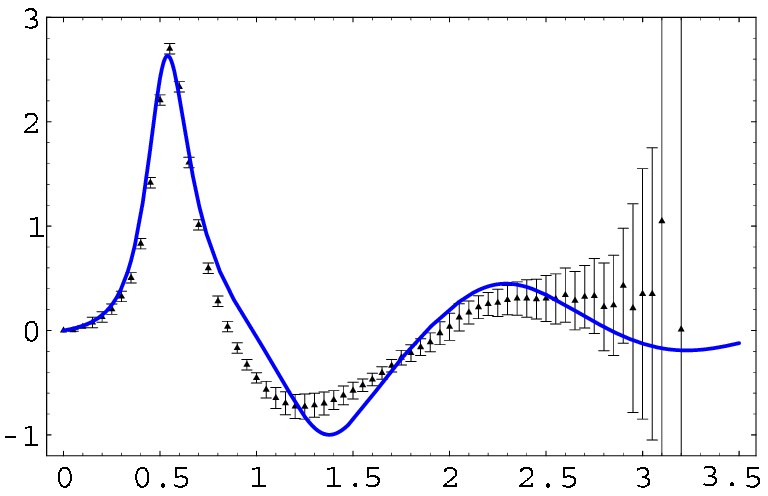}
\caption{Comparison of the data from ALEPH and OPAL (overlaid on a single plot and
including the kinematical factors of $\tau$ decay) with the model,
$\frac{1}{\pi}\mathrm{Im}~\Pi(q^2)$ from Eq.~(\ref{fifteencompact}) (solid blue
line), as a function of $q^2$ (in $\mathrm{GeV^2}$). As it is the case with the
experimental data, the pion contribution, $F_0^2/z$ in Eq.~(\ref{fifteencompact})
has been subtracted away.}\label{fig3}
\end{figure}

The choice of values for the parameters in Eq.~(\ref{nature}) plus the choice
\begin{equation}\label{aa}
    a=0.72
\end{equation}
for the new parameter $a$ produces the
spectral function which is plotted in Fig.~\ref{fig3}. As can be seen in this figure,
the model is not able to reproduce the experimental data in detail, but the
qualitative agreement is good enough for our purposes. We remind the reader that our
model is not meant to extract physical parameters, but only as a
laboratory for testing different methods which are currently in use for the
determination of the coefficients of the OPE from the data.

In the limit $N_c\rightarrow \infty$ one recovers for $\mathrm{Im}~\Pi(q^2)$ the
result with Dirac deltas of the previous section, Eq.~(\ref{spectrum}). However, at
small but finite $a/N_c$, the dependence of every resonance propagator on $q^2$
through the new variable $z$ produces a complex pole with an imaginary part, and
thus a
width $\Gamma_{V,A}(n)$ given by
\begin{equation}\label{seventeen}
    \Gamma_{V,A}(n)=\frac{a}{N_c} M_{V,A}(n)\ \left(1+ \mathcal{O}\left(\frac{a}{N_c}\right)\right) \ ,
\end{equation}
and, mutatis mutandis, for the $\rho$ as well. This behavior for $\Gamma$ is the one
expected on general grounds \cite{Shifman} and, as one can see, forces
$\Gamma_{V,A}(n)$ to vanish as $N_c\rightarrow \infty$, although it should be
noted that the width also grows with the resonance number $n$.

Given that the change produced by the widths amounts to the replacement
(\ref{sixteen}), the expansion of the Green's function $\Pi(q^2)$ at large values of
the momentum $|q^2|$ is similar to that in Eq.~(\ref{DVinf}):
\begin{equation}\label{DV}
\Pi(q^2)\approx \Bigg\{
\begin{array}{ll}
    \!\!\!\Pi_{OPE}(q^2) +\mathcal{O}\left(e^{-2\pi|q^2/\Lambda^2|^{\zeta}}\right)& ,
    \  \  \hbox{${\rm Re}\ q^2\leq0$\ ,} \\
    \!\!\!\Pi_{OPE}(q^2)+\Delta(q^2)+\mathcal{O}\left(e^{-2\pi|q^2/\Lambda^2|^{\zeta}}\right)&,\  \  \hbox{${\rm Re}~q^2\geq0$\ ,} \\
\end{array}
\end{equation}
except that now
\begin{equation}\label{deltaa}
    \Delta(q^2)=\frac{\pi F^2}{\Lambda^2}\ \frac{1}{\zeta}\ \Bigg\{\cot
    \left[\pi\left(\frac{-q^2}{\Lambda^2}\right)^{\zeta}+\pi \frac{m_V^2}{\Lambda^2}
    \right]- \cot \left[\pi\left(\frac{-q^2}{\Lambda^2}\right)^{\zeta}+
    \pi \frac{m_A^2}{\Lambda^2}\right]\!\Bigg\}\ ,
\end{equation}
with $\zeta=1-\frac{a}{\pi Nc}$. Similarly, the OPE is given by an expansion as in
Eq.~(\ref{three}) but the coefficients (\ref{ope}) are shifted by a $\log
Q^2$-dependent term. For small $a/N_c$, the Wilson coefficients read
\begin{equation}\label{eighteen}
    C_{2k}(Q^2)^{N_c=3} = C_{2k}^{N_c=\infty}
    \left(1+\frac{a}{\pi N_c}+ \frac{ka}{\pi N_c}\log\frac{Q^2}{\Lambda^2}
    +\mathcal{O}\left(\frac{a^2}{N_c^2}\right)\right)\ ,
\end{equation}
where the $C_{2k}^{N_c=\infty}$ are those given in Eq.~(\ref{ope}).  We will find it
convenient to explicitly separate the log dependence and decompose these
coefficients as \cite{Cirigliano}
\begin{equation}\label{ab}
     C_{2k}(Q^2)^{N_c=3}= a_{2k}+b_{2k}\
     \log\frac{Q^2}{\Lambda^2}+ \mathcal{O}\left(\frac{a^2}{N_c^2}\right)\ .
\end{equation}
The $b_{2k}$ would model the existence of anomalous dimensions in real QCD
were it not for the fact that the latter, unlike the $b_{2k}$, do not vanish in the
large-$N_c$ limit. We do not consider this to be  a serious drawback of the model,
however, and the $1/N_c$ contributions in Eq.~(\ref{eighteen}) still mimic
the $\alpha_s(m_{\tau})$ contributions to the anomalous dimensions which exist in
real QCD.
Even though the $1/N_c$ corrections in Eq.~(\ref{eighteen}) grow larger
with increasing dimension of the operator, at least for the first terms there is a
sense in which the $N_c\rightarrow\infty$ limit is close to the world at $N_c=3$ in
the OPE. Plugging in our set of numbers (\ref{nature}, \ref{aa}), the OPE
coefficients are
\begin{eqnarray}\label{opewidths}
a_2=b_2=&0&=a_4=b_4\nonumber\\
a_6= -2.8 \times 10^{-3} \mathrm{GeV}^6\quad &,&
\quad b_6=-5.9 \times 10^{-4} \mathrm{GeV}^6\ ,\nonumber\\
a_8= +1.8 \times 10^{-3} \mathrm{GeV}^8 \quad &,&\quad b_8= +5.1 \times
10^{-4}\mathrm{GeV}^8\ .
\end{eqnarray}

The small $1/N_c$ correction to the Green's function $\Pi(q^2)$ in the Euclidean
(cf. Eq.~(\ref{eighteen})) is in complete contradistinction to what happens to the
spectral function $\mathrm{Im}~\Pi(t)$ in the Minkowski region, $t>0$. As emphasized
in Ref.~\cite{Shifman}, the reason for this is that the limits
$N_c\rightarrow\infty$ and $t\rightarrow \infty$ do not commute in this region.
Defining $q^2=t>0$, one obtains to leading order in $a/N_c$ for the variable $z$
of Eq.~(\ref{sixteen}) in this region,
\begin{equation}\label{imaginary}
z = -t\left(1-\frac{a}{\pi
N_c}\log{\left(\frac{t}{\Lambda^2}\right)}+i\frac{a}{N_c}+
\mathcal{O}\left(\frac{a^2}{N_c^2}\right)\right)\ .
\end{equation}
Taking imaginary parts of the reflection property for the $\psi$ function in
Eq.~(\ref{reflexion}), at large values of the momentum $t>0$ one obtains that
\begin{equation}\label{oscillation}
    \mathrm{Im}\ \Pi(t)= \ \mathrm{Im}\ \Pi_{OPE}(t)\  +\
    \mathrm{Im}\ \Delta(t)\ ,
\end{equation}
where
\begin{equation}\label{imope}
    \mathrm{Im}\ \Pi_{OPE}(t)= \frac{3 a }{N_c}\ \frac{b_6}{t^3}
    \left(1+ \mathcal{O}\left(\frac{a}{N_c}\right)\right)+
    \mathcal{O}\left(\frac{1}{t^4}\right)\ ,
\end{equation}
and
\begin{eqnarray}\label{imdv}
\mathrm{Im}\ \Delta(t)\!&\!=\!&\!
\frac{4\pi F^2}{\Lambda^2}\ e^{-\frac{2\pi a}{N_c}\frac{t}{
\Lambda^2 }}\ \sin\left(\pi \frac{2t-m_A^2-m_V^2}{\Lambda^2}\right) \sin\left(\pi
\frac{m_V^2-m_A^2}{\Lambda^2}\right)\left(1+\mathcal{O}\left(\frac{a}{N_c}\right)
\right) \nn
\\ && +\  \mathcal{O}\left(e^{-\frac{4\pi a}{N_c}\frac{t}{ \Lambda^2 }}\right)\ .
\end{eqnarray}
The imaginary part in Eq.~(\ref{imope}) can be gotten from the imaginary part of the
logarithm in the OPE expansion in Eq.~(\ref{eighteen}) and is also familiar from the
QCD case, except for the $1/N_c$ suppression. Analogously, the cancelation of the
$t^{-1}$ and $t^{-2}$ terms is due to the absence of the corresponding $1/Q^2$ and
$1/Q^4$ terms in the OPE in the Euclidean region.

The contribution shown in Eq.~(\ref{imdv}) is completely missed by the
OPE and stems from the duality-violating function $\Delta(q^2)$ in
Eq.~(\ref{deltaa}). We call the reader's attention to
the $N_c$ factor in the denominator of the exponent in
Eq.~(\ref{imdv}). The expression in  this equation  is obtained in the large $t$
limit when $N_c$ is kept finite. That is to say, for finite $N_c$, the isolated
poles of $\Delta_{\infty}(q^2)$ on the real axis (Fig.~\ref{YYY}) become a cut. If,
on the contrary, the limit $N_c\rightarrow\infty$ is taken first, then the
expression (\ref{imdv}) is not valid and one obtains the infinite sum of delta
functions of the previous section instead.  It follows  that the two
limits $N_c\rightarrow\infty$ and $t\rightarrow \infty$ do not commute in the
Minkowski region. We expect this effect  to be quite generic, and we thus believe the exponentially
damped oscillation shown in Eq.~(\ref{imdv}) to be more general than our particular
model \cite{Shifman}.

In general, although
Eqs.~(\ref{ddelta},\ref{dualviol}) are still valid, the behavior of the duality-violating
function $\Delta(q^2)$ changes drastically with respect to its counterpart
$\Delta_\infty(q^2)$, showing an exponential fall-off at large $|q^2|$ on the half
plane $\mathrm{Re}~q^2\geq 0$. One has now that
\begin{equation}\label{falloff}
    \Delta(q^2)\ \sim_{_{_{\!\!\!\!\!\!\!\!\!\! \!\!\!|q^2|\ \mathrm{large}}}}
    e^{-2\pi\left(\frac{|q^2|}{\Lambda^2}\right)^{\zeta}\  \big|\sin\big\{\zeta
    (\varphi-\pi)\big\}\big|}\ ,\quad q^2=|q^2|\ e^{i\varphi}\ ,\quad
\left\{%
\begin{array}{ll}
     0\leq \varphi\leq
    \frac{\pi}{2}\\
    \frac{3 \pi}{2}\leq \varphi\leq 2\pi \\
\end{array}%
\right. \ .
\end{equation}
We see in particular that the limit $N_c\rightarrow\infty$ (i.e. $\zeta
\rightarrow 1$) makes the values $\varphi=0,\ 2\pi$ exceptional: when  $N_c=\infty$,
the exponential fall-off on the real axis completely disappears. This is of course
consistent with Eq.~(\ref{imdv}).

The exponential fall-off exhibited in Eq.~(\ref{falloff}) implies now that the
function $\mathcal{D}^{[n]}(s_0)$ vanishes exponentially at large $s_0$ as
\begin{equation}\label{falloff2}
    \mathcal{D}^{[n]}(s_0) \sim \mathcal{O}
    \left(e^{-\frac{2 \pi a}{N_c} \frac{s_0}{\Lambda^2}}\right)\ .
\end{equation}
Again we note the singular limit
$N_c\rightarrow \infty$ in this expression: in the
previous section we saw that $\mathcal{D}^{[n]}(s_0) \rightarrow $ ``undamped
oscillations," if the limit $N_c\rightarrow \infty$ is taken before $s_0$ is taken
to infinity.

For finite $N_c$, one has
\begin{eqnarray}\label{a}
    \mathcal{D}^{[n]}(s_0)&=&  \mathcal{D}_{oscill.}^{[n]}(s_0)+
    \mathcal{D}_{const.}^{[n]}(s_0)\qquad
    \sim_{_{\!\!\!\!\!\!\!\!\!\!\!s_0\,\mathrm{large}}} \quad \mathcal{O}
    \left(e^{-\frac{2 \pi a}{N_c}\frac{ s_0}{\Lambda^2}}\right)\ ,\\
 \mathcal{D}_{oscill.}^{[n]}(s_0)&=&\int_{0}^{s_0} dt\ t^n\ \frac{1}{\pi}
    \ \mathrm{Im}\ \Delta(t+i\varepsilon)\qquad
    \sim_{_{\!\!\!\!\!\!\!\!\!\!\!\!s_0\,\mathrm{large}}} \ -\mathcal{C}^{[n]} +
    \mathcal{O}\left(e^{-\frac{2 \pi a}{N_c}\frac{s_0}{\Lambda^2}}\right)\label{b}\ ,\\
   \mathcal{D}_{const.}^{[n]}(s_0) &=&\ \mathcal{C}^{[n]}\ +\
      \mathcal{O}\left(e^{-2 \pi \frac{s_0}{\Lambda^2}}\right)\ .\label{c}
\end{eqnarray}
We are using the hierarchy
\begin{equation}\label{hierarchy}
    e^{-\frac{2 \pi a }{N_c}\frac{s_0}{\Lambda^2}}\gg e^{-2 \pi
    \frac{s_0}{\Lambda^2}}\ ,
\end{equation}
which is very well satisfied  for large values of $s_0$.
Note the more stringent bound on the corrections to Eq.~(\ref{c}), which results from
the fact that $\mathcal{D}_{const.}^{[n]}(s_0)$ is defined as an integral over
$\Delta(q^2)$ for values of $q^2$ far away from the positive real axis.%
\footnote{See the Appendix for the derivation of Eq.~(\ref{c}).}

Combining Eqs.~(\ref{a}-\ref{c})  in the $s_0\rightarrow \infty$ limit, it follows
that
\begin{equation}\label{d}
    \mathcal{C}^{[n]}=- \int_{0}^{\infty} dt\ t^n\ \frac{1}{\pi}
    \ \mathrm{Im}\ \Delta(t+i\varepsilon)\ ,
\end{equation}
and, inserting this result into Eq.~(\ref{a}), we obtain
\begin{equation}\label{finally}
    \mathcal{D}^{[n]}(s_0)=- \int_{s_0}^{\infty} dt\ t^n\ \frac{1}{\pi}
    \ \mathrm{Im}\ \Delta(t+i\varepsilon)\ +\
    \mathcal{O}\left(e^{-\frac{2 \pi s_0}{\Lambda^2}}\right)
\end{equation}
as the final expression in the finite width case, for the duality violations
in Eq.~(\ref{ddelta}). Since
$\mathrm{Im}~\Delta(t+i\varepsilon) \sim \mathcal{O}(e^{-\frac{2 \pi
a}{N_c}\frac{t}{\Lambda^2}})$ (cf. Eq.~(\ref{imdv})), the integral in
Eq.~(\ref{finally}) is indeed of order $e^{-\frac{2 \pi s_0}{N_c \Lambda^2}}$, in
agreement with Eq.~(\ref{a}).

\begin{figure}
\renewcommand{\captionfont}{\small \it}
\renewcommand{\captionlabelfont}{\small \it}
\centering
\includegraphics[width=2.5in]{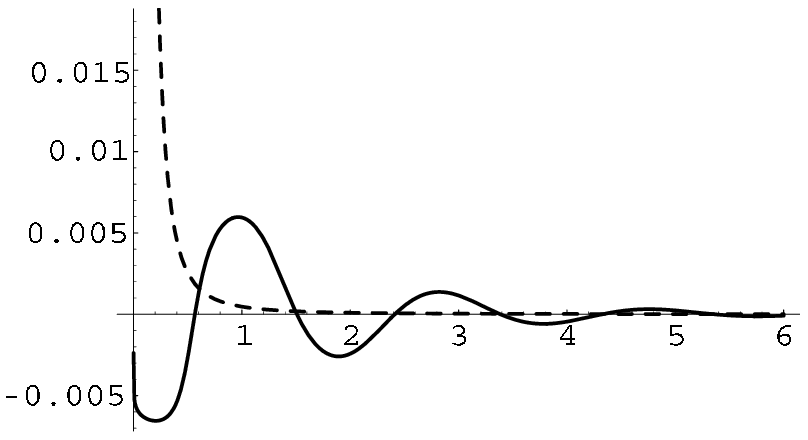}
\hspace {1cm}
\includegraphics[width=2.5in]{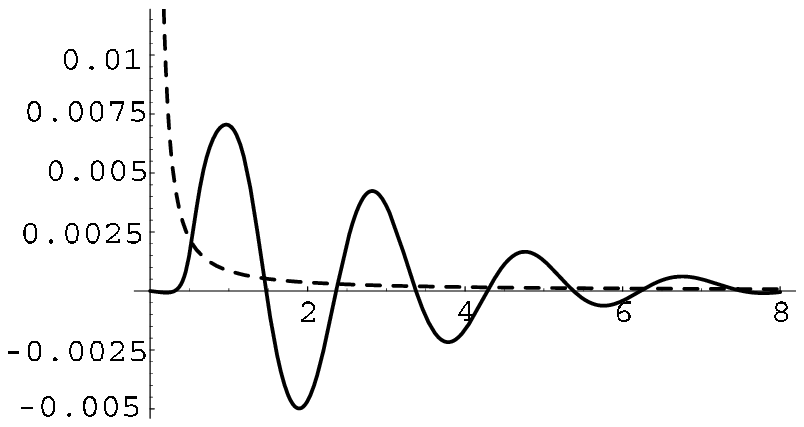}\\
\vspace {1cm}
\includegraphics[width=2.5in]{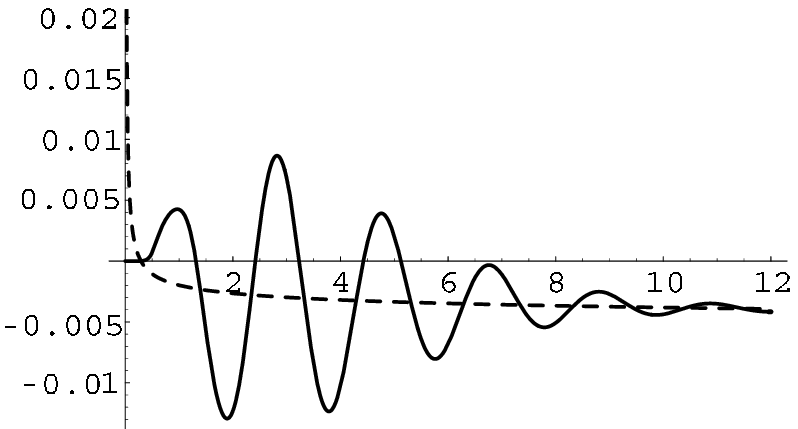}
\hspace{1cm}
\includegraphics[width=2.5in]{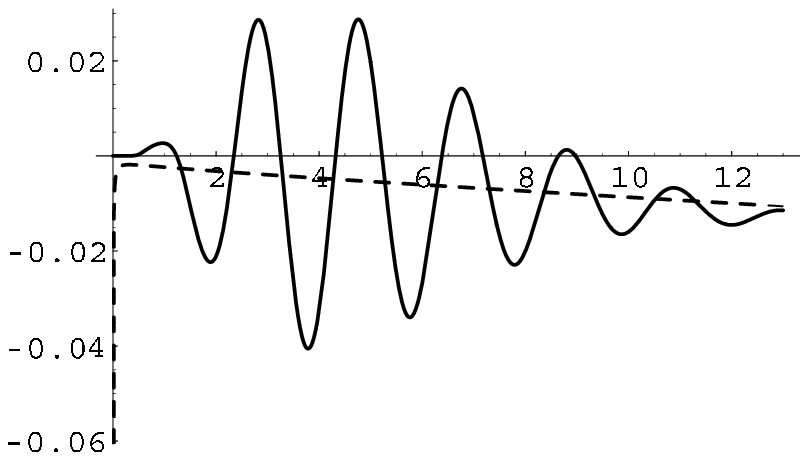}
\caption{From left to right: $M_0$, $M_1$ (solid curves, first row), $M_2$ and $M_3$
(solid curves, second row), defined in Eq. (\ref{five}), together with the OPE
prediction (dashed curves) as given by the corresponding terms in squared brackets
in Eqs. (\ref{works0}-\ref{works3}), as functions of the energy scale, $s_0$ (in
$\mathrm{GeV}^2$).}\label{MMM}
\end{figure}

Returning to Eq.~(\ref{ddelta}), we obtain for the OPE coefficients the
following set of equations:
\begin{eqnarray}\label{works0}
 \int_{0}^{s_0}\!\!\!\!\!dt\ \rho(t)+\left[\frac{b_6}{2 s_0^2}-\frac{b_8}{3s_0^3}+...  \right]
  \!\!\!\!&=&\!\!\!  \mathcal{D}^{[0]}(s_0) \\\label{works1}
  \int_{0}^{s_0}\!\!\!\!\!dt\ t\ \rho(t)+\left[\frac{b_6}{s_0}-\frac{b_8}{2 s_0^2}+...  \right]
  \!\!\!&=&\!\!\!  \mathcal{D}^{[1]}(s_0)  \\\label{works2}
  \int_{0}^{s_0}\!\!\!\!\!dt\ t^2\rho(t)-\left[a_6+ b_6 \log\frac{s_0}{\Lambda^2}+
  \frac{b_8}{s_0}+...\right] \!\!\!&=& \!\!\!\mathcal{D}^{[2]}(s_0)
  \\\label{works3}
  \int_{0}^{s_0}\!\!\!\!\!dt\ t^3\rho(t)+\left[a_8+b_8 \log\frac{s_0}{\Lambda^2}-
   b_6 s_0 +  \frac{b_{10}}{s_0}+...  \right]\!\!\!&=& \!\!\!
   \mathcal{D}^{[3]}(s_0) \\
&\vdots & \nn \\
 \!\!\!\!\!\!\!\!\!\!\!\!\! \int_{0}^{s_0}\!\!\!\!\!dt\ t^{7}\rho(t)+\left[a_{16}+b_{16}
 \log\frac{s_0}{\Lambda^2}-
\frac{b_6 s_0^5}{5}+ \frac{b_8 s_0^4}{4}+\ldots+ \frac{b_{18}}{s_0}+...
\right]\!\!\! &=&\!\!\!
    \mathcal{D}^{[7]}(s_0)\ , \label{works5}
\end{eqnarray}
where the right-hand side is given by Eq.~(\ref{finally}). The results for the
lowest moments (\ref{five}) and the corresponding OPE (the terms explicitly shown in
squared brackets in Eqs. (\ref{works0}-\ref{works3})) are plotted in Fig. \ref{MMM}.

We draw
the following conclusions from Eqs.~(\ref{works0}-\ref{works5}):
First, it is clear that the $b$ coefficients of the OPE also contribute to these
equations. The pattern is simple and governed by dimensional analysis: a higher
order
$b$ contributes to the equation determining a lower order $a$
accompanied by inverse powers of the scale $s_0$.
On the other hand, a lower order $b$ contributes to the equation determining a
higher order $a$ accompanied by \emph{positive} powers of $s_0$.

Even though in real QCD the $b$ coefficients are order $\alpha_s$ effects with
respect to the $a$'s, the fact that they may appear multiplied by positive powers of
$s_0$ tells us that, in general, it is safer not to neglect them from the start,
unlike what is done at present in common practice. A possible exception is the term $b_6$, which is the
only one which has been calculated  \cite{Cirigliano2} and which
seems to be small enough to
be safely disregarded. Since this term is the only one accompanied by positive
powers of $s_0$ in Eqs.~(\ref{works0}-\ref{works3}), it might not be too
unreasonable to neglect the $b$ terms in these equations in a first approximation.
However, for higher moments, this practice may be dangerous. Notice that
an $s_0^4$ term amounts to a factor of almost 100 $\mathrm{GeV}^{8}$ for
$s_0=m_{\tau}^2=3.15~\mathrm{GeV}^2$ so that, for instance, the term $b_8 s_0^4$ in
the last Eq.~(\ref{works5}) might not be negligible even if $b_8$ is small.

Second, and most important, the right-hand sides of Eqs.~(\ref{works0}-\ref{works5})
do not vanish in general, except in the limit $s_0\rightarrow \infty$. As it is
obvious from these equations, the $\mathcal{D}$ terms may potentially pollute the
extraction of the OPE coefficients. Consequently, in any precise determination of
OPE coefficients it is unavoidable to take this duality violation into account.

In order to further appreciate this point,  let us close our eyes, neglect the
duality violations $\mathcal{D}$ altogether, and apply the different methods which
have been employed in the literature so far.

For instance, let us start with finite-energy sum rules. The method consists in
determining a duality point $s_0^*$ from the condition $M_{0,1}(s_0^*)\simeq 0$,
which is then to be used in $M_{2,3}(s_0^*)$ to predict the combinations appearing in
Eqs.~(\ref{works2},\ref{works3}), namely
\begin{eqnarray}\label{maarten}
    A_6(s_0^*)&\equiv& a_6+ b_6 \log \frac{s_0^*}{\Lambda^2}+ \frac{b_8}{s_0^*}+... \ , \nn \\
    A_8(s_0^*)&\equiv& a_8+ b_8 \log \frac{s_0^*}{\Lambda^2}- b_6 s_0^* +
    \frac{b_{10}}{s_0^*}+...
\end{eqnarray}
In our model the first duality point happens to sit at $ s_0^{*}=1.472
\,\,{\mathrm{GeV}}^2$, yielding the predictions
\begin{equation}\label{fesr1}
A_6^{FESR}=-4.9 \cdot 10^{-3} \,\,{\mathrm{GeV}}^6 \quad , \quad A_8^{FESR}=+9.3
\cdot 10^{-3} \,\, {\mathrm{GeV}}^8\ ,
\end{equation}
which can be compared to the correct values
\begin{equation}\label{truevalues}
     A_6(1.472 \ {\mathrm{GeV}}^2)= -2.4\cdot
10^{-3} \mathrm{GeV}^6 \quad , \quad A_8(1.472\ {\mathrm{GeV}}^2)= +2.6\cdot 10^{-3}
\mathrm{GeV}^8\ .
\end{equation}
There is a second duality point around $2.4\ \mathrm{GeV}^2$ fulfilling either
$M_0(s_0^*)=0$ or $M_1(s_0^*)=0$, but not both. Choosing the one which satisfies
$M_1(s_0^*)=0$, for instance, one gets $s_0^*=2.363\ \mathrm{GeV}^2$ and this yields
\begin{equation}\label{fesr2}
A_6^{FESR}=-2.0 \cdot 10^{-3} \,\,{\mathrm{GeV}}^6 \quad , \quad A_8^{FESR}=-1.6
\cdot 10^{-3} \,\, {\mathrm{GeV}}^8\ ,
\end{equation}
to be compared to
\begin{equation}\label{truevalues2}
A_6(2.363 \ {\mathrm{GeV}}^2)= -2.8\cdot 10^{-3} \mathrm{GeV}^6\quad , \quad
A_8(2.363\ {\mathrm{GeV}}^2)= +3.4\cdot 10^{-3} \mathrm{GeV}^8\ .
\end{equation}

As we see, in going from the first to the second duality point, $A_6$ is reduced by
a factor 2 and $A_8$ changes sign. Interestingly enough, this is precisely the trend
seen in the corresponding determinations in the literature \cite{Phily, Bijnens}.

{}From the comparison of both results (\ref{fesr1}) and (\ref{fesr2}), it seems that
there is no dramatic advantage to making the prediction using the second duality
point rather than the first, even though at the second point $s_0$ has a larger
value.\footnote{The prediction for $A_6$ at the second $s_0^*$ is somewhat better
than at the first, but this is no longer the case for $A_8$.} Moreover, while going
to a larger $s_0$ is in principle better because the duality violations
$\mathcal{D}^{[n]}(s_0)$ become smaller, if this is done while keeping all the $b$
coefficients to zero, i.e. taking the $A$'s above as estimates for the $a$'s, then
the growth of the moments $M_{2,3}(s_0)$ with $s_0$ --- they are actually divergent
because of Eq.~(\ref{imope}) --- is numerically absorbed by a wrong shift in
$a_{6,8}$. As we see in Eqs.~(\ref{fesr1}) and (\ref{fesr2}), this may invalidate
the advantage of choosing a larger $s_0$. Furthermore, in the real world the
experimental error bars are much bigger at larger $s_0$, the QCD value of $b_6$ is
very small and that of $b_8$ is unknown. This makes it difficult to know exactly
what explains the difference between the results obtained in the literature using
finite-energy sum rules \cite{Phily, Bijnens,
Narison}.\footnote{Reference~\cite{Rojo} finds numerical agreement with
Ref.~\cite{Bijnens} but, in fact, it does not have the second duality point the
analysis of Ref.~\cite{Bijnens} is based on.}

\begin{figure}
\renewcommand{\captionfont}{\small \it}
\renewcommand{\captionlabelfont}{\small \it}
\centering
\includegraphics[width=2.5in]{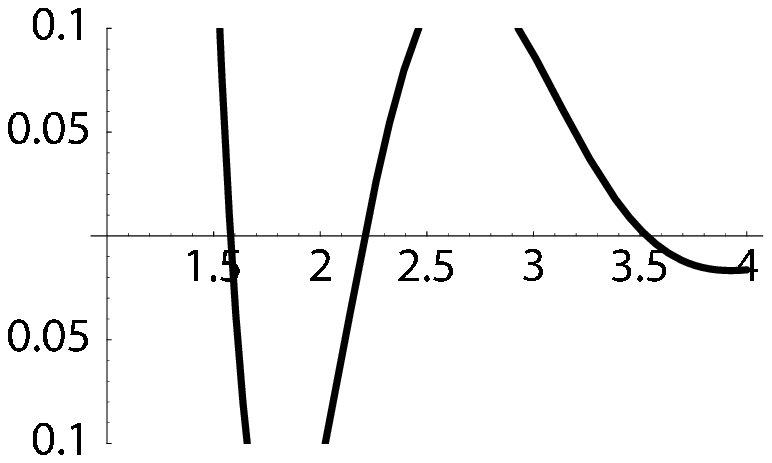}
\includegraphics[width=2.5in]{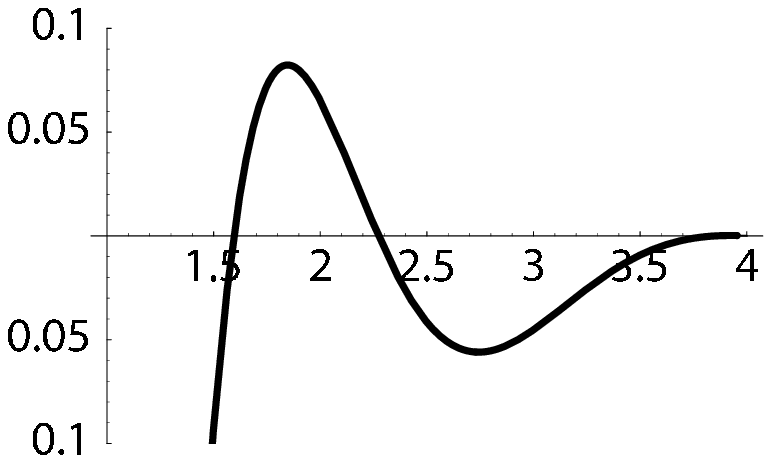}\\
\vspace {0.5cm}
\includegraphics[width=2.5in]{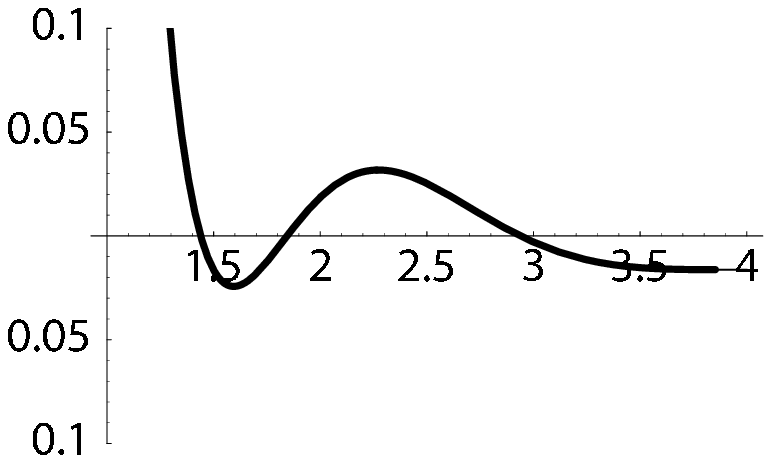}
\caption{Difference between the integrals of the pinched weights, $\int_0^{s_0}dt\
P_{1,2,3}(t)\rho(t)$ and the corresponding OPE contribution from
Eqs.~(\ref{works0}-\ref{works3}) (in units of $10^{-3}~{\rm GeV}^2$), as a function of $s_0$ (in
$\mathrm{GeV^2}$), from left to right and top to bottom, respectively.}\label{YY}
\end{figure}

Let us next consider pinched weights, which have been suggested \cite{Cirigliano, Dominguez} as a way to
reduce the amount of duality violations. However, the use of pinched weights is no
guarantee that one can do away with the duality violations altogether. For instance,
we have followed Ref.~\cite{Cirigliano} and fitted the OPE coefficients to the
combination of moments obtained with the pinched weights $P_{1}=(1-3
t/s_0)(1-t/s_0)^2$ and $P_{2}=(t/s_0)(1-t/s_0)^2$, neglecting duality violations. We
have done this with 20 points in the window $1.5 \ \mathrm{GeV}^2\leq s_0 \leq 3.5\
\mathrm{GeV}^2$. The upper end in this window was identified by noting that the tau
mass roughly coincides with the mass of the $\rho(1700)$ in QCD, and in our model
this resonance happens to be at $m_{\rho''}^2\simeq 3.8\ \mathrm{GeV}^2$. The result
from the fit can be interpreted as some sort of average over the window in $s_0$ of
the combinations $A_{6,8}$ in Eq.~(\ref{maarten}), and yields
\begin{equation}\label{pw}
    A_6^{pinch}= -3.8\times 10^{-3}\ \mathrm{GeV}^6 \quad ,
    \quad A_8^{pinch}= 6.5 \times 10^{-3}\ \mathrm{GeV}^8 \ .
\end{equation}
As one can see, the error made is again rather large (particularly in the case of
$A_8$) and, in fact, comparable to that using finite-energy sum rules in
Eqs.~(\ref{fesr1}, \ref{fesr2}). We did not see any clear improvement in precision by
changing details of the fit such as, e.g., the window of $s_0$ values used.

Since pinched weights suppress duality violations, one may try to design ``good''
pinched weights which have as small duality violations as possible. Assuming a
general expression for the duality violations of the form (cf. Eq.~(\ref{imdv}))
\begin{equation}\label{fit}
    \frac{1}{\pi} \mathrm{Im}\Delta(t)= \kappa\  e^{-\gamma t}\  \sin\left(\alpha +
    \beta t\right)\ ,
\end{equation}
for certain values of the parameters $\kappa, \gamma, \alpha$ and $\beta$, one could
ask which is the pinched weight, involving moments not higher than $M_3$, which
minimizes the amount of duality violations. As before, the answer to this question is
$P_3=(1-t/s_0)^3$. The reason is that, given the general form in Eq.~(\ref{fit}),
one finds that $P_3$'s residual oscillation is damped by the factor $s_0^{-3}
e^{-\gamma s_0}$, while in the case of $P_{1,2}$ the damping factor is only
$s_0^{-2} e^{-\gamma s_0}$. This is a behavior reminiscent of the  $N_c=\infty$ case
in the previous section.\footnote{This is a general result. For moments not higher
than $M_n$ the optimal pinched weight is $(1-t/s_0)^n$ and the residual oscillation
is modulated by $s_0^{-n} e^{-\gamma s_0}$.} Figure~\ref{YY} shows the difference
between the integral of the pinched weights $P_{1,2,3}$ and their corresponding OPE
contribution from Eqs.~(\ref{works0}-\ref{works3}) as a function of $s_0$. If there
were no duality violations these curves should be a flat zero. As this figure shows,
$P_3$ has the smallest duality violations.

\begin{figure}
\renewcommand{\captionfont}{\small \it}
\renewcommand{\captionlabelfont}{\small \it}
\centering
\includegraphics[width=2.5in]{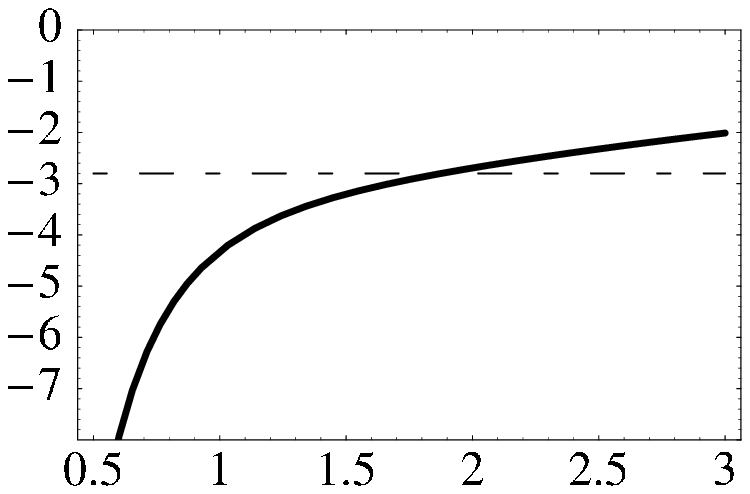}
\hspace {1cm}
\includegraphics[width=2.4in]{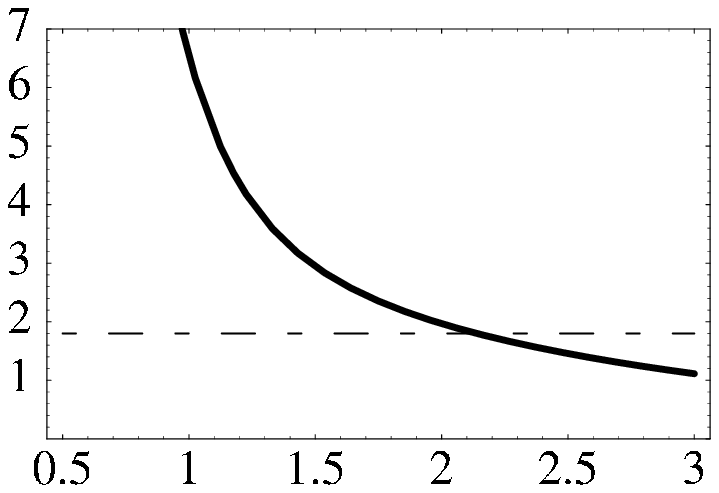}\\
\caption{Plot of the coefficients $a_6^L$ (solid curve, left) and $a_8^L$ (solid
curve, right), together with the corresponding true values $a_{6,8}$ (dashed
horizontal lines) from Eqs.~(\ref{opewidths}) (all in units of $10^{-3}$), as a
function of the Laplace variable, $\tau$ (in $\mathrm{GeV^{-2}}$).  }\label{fig7}
\end{figure}

In Ref.~\cite{Narison} a Laplace sum rules analysis was done. Neglecting the $b$
coefficients one obtains for the $a$ coefficients of the OPE  the equations
\begin{eqnarray}\label{laplace}
a_6^L - a_{10}^L\ \frac{\tau^2}{12} +...&=& \frac{6}{\tau^2} \int_0^{s_0}dt\ e^{-t
\tau}\ \rho(t)+
\frac{2}{\tau} \int_0^{s_0}dt\ t\ e^{-t \tau}\ \rho(t)\ ,\nn \\
a_8^L+ a_{10}^L\ \frac{\tau}{2} +...&=& -\frac{12}{\tau^3} \int_0^{s_0}dt\ e^{-t
\tau}\ \rho(t)- \frac{6}{\tau^2} \int_0^{s_0}dt\ t\ e^{-t \tau}\ \rho(t) \ .
\end{eqnarray}
 Requiring stability under variations in the Laplace variable $\tau$
 one can determine $a_{6,8}^L$.\footnote{
 Obviously, the true $a$ coefficients (\ref{opewidths}) are independent of
$\tau$.} In Fig.~\ref{fig7} we have plotted the true values for $a_{6,8}$ from
Eq.~(\ref{opewidths}) together with the $\tau$ dependence of $a_6^L$, $a_8^L$ --
neglecting $a_{10}^L$ and higher. The upper limit in the Laplace integral has been
chosen to be the second duality point from the previous finite-energy sum rules
analysis of Eq.~(\ref{fesr2}), $s_0^*\simeq 2.35\ \mathrm{GeV}^2$, but there is no
qualitative change if the first duality point (\ref{fesr1}) is used instead. As it
can be clearly seen, no sign of stability in $\tau$ is found.   (We do not expect
this to change qualitatively if $a_{10}$, $b_6$ etc. are included.)  This is unlike
what is found in QCD \cite{Narison} and may be another sign of what we already said
in the introduction concerning the model's tendency to maximize the violations of
duality.

Finally, we also used the MHA method \cite{MHA}. In this case one truncates the
spectrum to a finite number of resonances whose parameters are adjusted so that
the low- and high-energy expansions of  the $\Pi(q^2)$ function in the large-$N_c$
limit are reproduced. In this way one constructs an interpolator for the function
$\Pi(q^2)$, called $\Pi_{MHA}(q^2)$ with which it is possible to calculate, e.g.,
the integrals which determine the relevant couplings governing semileptonic kaon
decays. However, in the present case the situation is slightly different because one
wishes to use $\Pi_{MHA}(q^2)$ to predict coefficients of the high-energy
expansion, and not as an interpolator for an integrand.

The function $\Pi_{MHA}(q^2)$ is constructed as\footnote{See Ref.~\cite{Friot} for a
more elaborated version of this method applied to QCD, in which also a $\rho'$ is
included.}
\begin{equation}\label{MHA}
    \Pi (q^2)^{MHA}=\frac{F_{0}^2}{q^2}+
\frac{{\widetilde{F}_V}^2}{{\widetilde{M}_{V}}^2-q^2}-
\frac{{\widetilde{F}_{A}}^2}{{\widetilde{M}_{A}}^2-q^2} \ ,
\end{equation}
with the parameters $\widetilde{M}_V, \widetilde{M}_A, \widetilde{F}_V,
\widetilde{F}_A$ to be determined from the conditions $C_{2,4}=0$ and the values of
$L_{10}$ and the pion electromagnetic mass difference in the large-$N_c$ limit, as
taken from the model in the previous section,  Eq. (\ref{onecompact}).

They are given by the following expressions:
\begin{eqnarray}\label{constraints}
    L_{10}&=& -\frac{1}{4}F_{0}^2\frac{\widetilde{M}_{A}^2+
\widetilde{M}_{V}^2}{\widetilde{M}_{V}^2 \widetilde{M}_{A}^2}=-(5.2\pm 0.5) \cdot 10^{-3}\\
\!\!\!\!m_{\pi}^{+}-m_{\pi}^{0}\!\!\!\!&=&\!\!\!\!\! \left(\frac{3\alpha}{8\pi^2
m_{\pi}F_{\pi}^2}\right)F_{0}^2 \frac{\widetilde{M}_{A}^2
\widetilde{M}_{V}^2}{\widetilde{M}_{A}^2-
\widetilde{M}_{V}^2}\log\left(\frac{\widetilde{M}_{A}^2}{\widetilde{M}_{V}^2}\right)=
(4.2\pm 0.4)\cdot 10^{-3}\,\,{\mathrm{GeV}}\ ,\nonumber
\end{eqnarray}
where $\alpha\simeq 1/137$ is the electromagnetic coupling constant, and the error
has been estimated on account of the large-$N_c$ approximation of the model.
Equations~(\ref{constraints}) fix the parameters in Eq.~(\ref{MHA}) to be
    \begin{eqnarray}\label{solMHA}
\widetilde{M}_{V}=0.70\pm 0.01\,\,{\mathrm{GeV}}\quad &,& \quad
\widetilde{M}_{A}=1.00\pm0.03\,\,{\mathrm{GeV}}\nonumber\\
\widetilde{F}_{V}=122\pm 6\,\,{\mathrm{MeV}}\quad &,& \quad \widetilde{F}_{A}=84\pm
7 \,\,{\mathrm{MeV}}\ .
\end{eqnarray}
Feeding expression (\ref{MHA}) with these values one has a prediction for the full
vacuum polarization function. In the Euclidean region the prediction is very good
for a broad range of intermediate values of $q^2$. For example, at $q^2=- 1\
\mathrm{GeV}^2$ the error between $\Pi(q^2)$ and $\Pi_{MHA}(q^2)$ is only $\sim 3\%$
\footnote{See also Ref. \cite{Philyandco}.}.

Expanding Eq.~(\ref{MHA}) in inverse powers of $q^2$, we obtain the following
prediction for the OPE coefficients:
\begin{equation}\label{numbersMHA}
a_6^{MHA}=- (3.6\pm 0.3) \times 10^{-3}\,\,{\mathrm{GeV}}^6\quad , \quad
a_8^{MHA}=(5.4 \pm0.7) \times 10^{-3}\,\,{\mathrm{GeV}}^8\ ,
\end{equation}
to be compared to the results in Eq.~(\ref{opewidths}). Again, no dramatic
improvement is found as compared to previous methods. As with other methods, the
prediction for $a_8$ is worse than that for $a_6$.

 In summary, the errors in the determinations of OPE coefficients from the
 different methods tend to be large in the present model. No method can be claimed
 to be more precise  than the others, and the error made is comparable in size
to the spread of values among the different methods.

It is difficult to say what happens in the case of QCD.  For $a_6$, the spread of
values in QCD found in the literature is not as bad as in the case of our model.
This could be related to the fact that the size of the coefficient $b_6$ relative to
$a_6$ is much smaller in QCD than it is in the model. But also in QCD the
discrepancies are clearly more serious for $a_8$ and higher coefficients
\cite{Narison}.

One of the lessons from our analysis of the model is that one should not ignore
duality violations represented by the $\mathcal{D}$ terms in
Eqs.~(\ref{works0}-\ref{works5}). In Fig.~9, we have plotted these $\mathcal{D}$
terms (solid lines). As one can see, they show zeros at certain values of $s_0\sim
1.5, 2.4, 3.4, ...\ $ GeV. These zeros correspond to values of $s_0$ where the
duality violations vanish and they are therefore the optimal points which one would
like to use as duality points in any finite-energy sum rule analysis. However, the
location of these zeros is not precisely determined because, as Fig.~9 shows, the
zeros are not universal but move from one spectral moment to the next. This is one
of the reasons why the finite-energy sum rule determination of OPE coefficients in
Eqs.~(\ref{maarten}-\ref{truevalues2}) has large errors. Consequently, as such
duality violations are likely to be present in QCD as well, we will now use the
model to show how these $\mathcal{D}$ terms may be included in a realistic analysis
also in QCD.

Our basic assumption will be that our result (\ref{finally}) together with the
functional form of a damped oscillation as in Eq.~(\ref{imdv}) are rather generic
results which go beyond our particular model. This assumption seems
reasonable\footnote{See also Ref.~\cite{Shifman}.} as these general features
basically depend on large $N_c$ and on the hierarchy of exponentials
(\ref{hierarchy}) controlling the different contributions in
Eqs.~(\ref{falloff2}-\ref{c}). In turn,
this depends only on the fact that large-$N_c$
duality violations are concentrated on the real axis in the complex $q^2$ plane.

Therefore, let us start by assuming a duality violation of the form given in
Eq.~(\ref{fit}), with parameters $\kappa, \gamma, \alpha$ and $\beta$ to be determined.
We remind the reader that in Eqs.~(\ref{works0},\ref{works1}) the $b$ terms are
suppressed by inverse powers of $s_0$. Therefore, in a first approximation, it may
be reasonable to completely neglect the $b$ terms in these two equations and,
through Eq.~(\ref{finally}), extract these parameters $\kappa, \gamma, \alpha$ and
$\beta$ by means of a simultaneous fit to Eqs.~(\ref{works0},\ref{works1}). Then the
rest of coefficients $a_{6,8,...}$ and $b_{6,8,...}$ can be obtained from
Eqs.~(\ref{works2}-\ref{works5}) in a straightforward way.

\begin{figure}
\renewcommand{\captionfont}{\small \it}
\renewcommand{\captionlabelfont}{\small \it}
\centering
\includegraphics[width=2.3in]{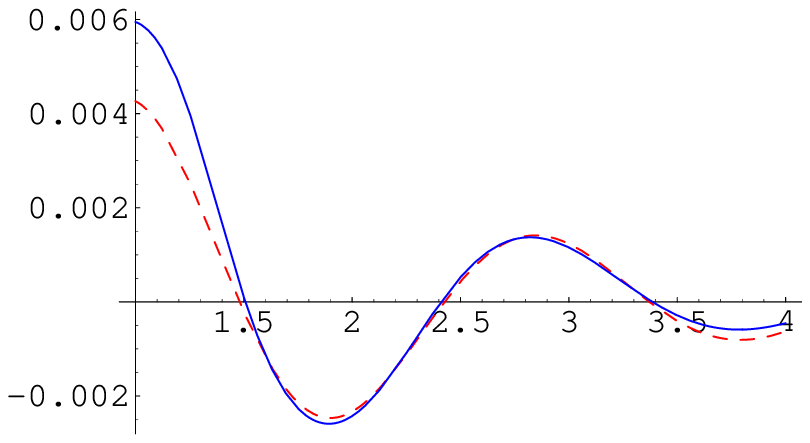}
\hspace{1cm}
\includegraphics[width=2.3in]{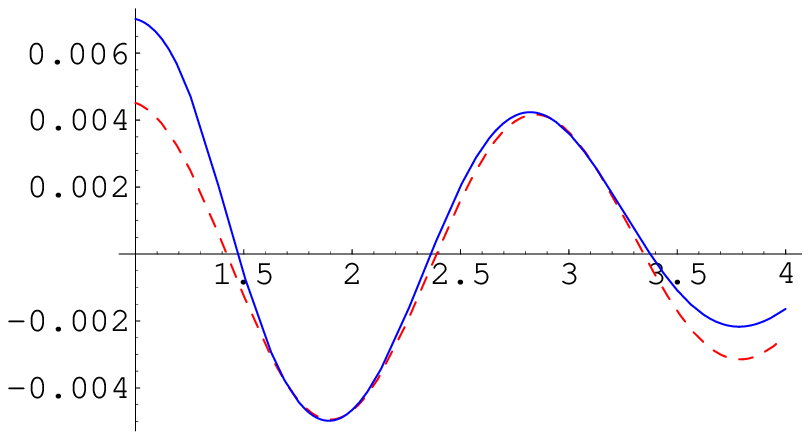}
\includegraphics[width=2.3in]{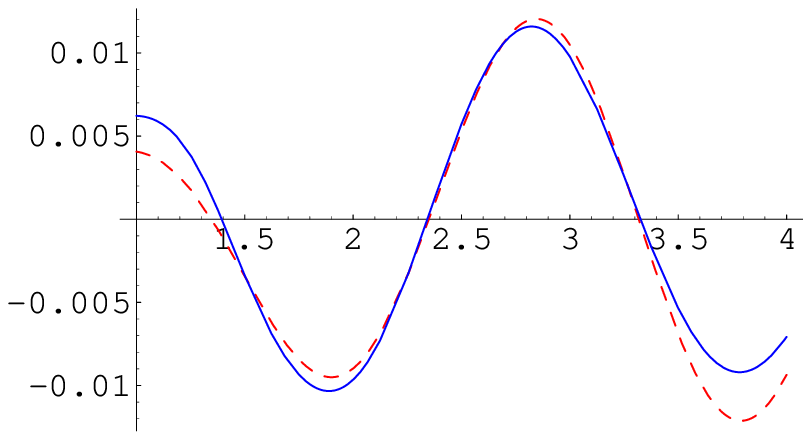}
\hspace{1cm}
\includegraphics[width=2.3in]{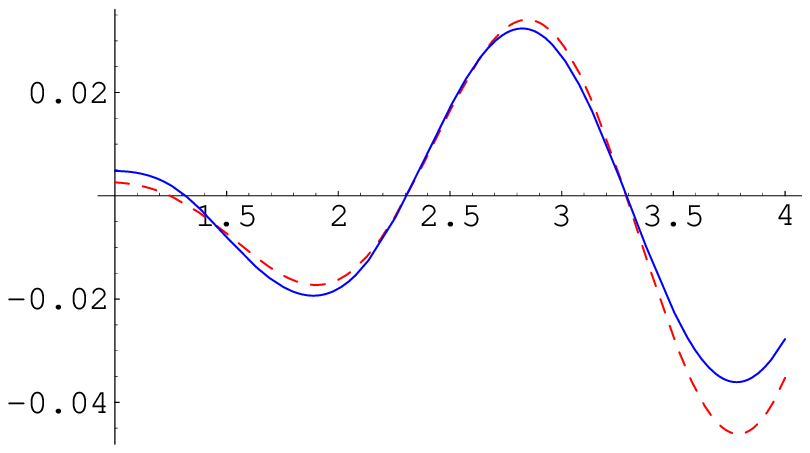}
\caption{Upper row: Plot of the left-hand side (solid line) and fit of the
right-hand side (dashed line) of Eqs.~(\ref{works0},\ref{works1}), neglecting the $b$
coefficients, as a function of $s_0$ (in $\mathrm{GeV^2}$). Lower row: Plot of the
left-hand side (solid line) and right-hand side predicted by the fit (dashed line)
of Eqs.~(\ref{works2},\ref{works3}), as a function of $s_0$ (in
$\mathrm{GeV^2}$).}\label{ZZZ}
\end{figure}

Just to illustrate how this would work in our model, we have determined the
parameters in Eq.~(\ref{fit}) by doing this simultaneous fit in the window $1.5\
\mathrm{GeV}^2\leq s_0 \leq 3.5\ \mathrm{GeV}^2$ with 20 equally-spaced points, with
the following result:
\begin{equation}\label{resultfit}
\kappa = 0.026\quad ,\quad \gamma= 0.591\ \mathrm{GeV}^{-2}\quad ,\quad \alpha=
3.323\quad \mathrm{and} \quad  \beta= 3.112\ \mathrm{GeV}^{-2}\ .
\end{equation}
Fig.~\ref{ZZZ} shows in the upper row the result of the fit to
Eqs.~(\ref{works0},\ref{works1}). As one can see, the overall fit is quite good. And
since it improves in the intermediate region of the fitting window where both
$\mathcal{D}^{[0]}$ and $\mathcal{D}^{[1]}$ have a zero, this suggests that a good
strategy is to determine the OPE at these zeros. This is what we will do next.

In the lower row of Fig.~\ref{ZZZ}, the dashed curve is the result for
$\mathcal{D}^{[2,3]}(s_0)$ from the fitted function (\ref{fit}) and the solid curve
is the result for the left-hand side of Eqs.~(\ref{works2},\ref{works3}), as
calculated from the true values of the model.

Armed with Eq.~(\ref{fit}) and the values of the parameters from the fit
(\ref{resultfit}), we predict a duality point $s_0^*$ at which
$\mathcal{D}^{[2]}(s_0^*)=0$,  within the window of our fit, to be at $s_0^*= 2.350\
\mathrm{GeV}^2$. Although similar, this value is not quite the same as the one
obtained in the finite-energy sum rules analysis before, which was $2.363$ GeV$^2$.
This small difference has a large impact on the determination of the OPE due to the
steepness of the slope in the plot of Fig. 9 (lower curves), resulting in the large
error quoted in the finite-energy sum rule analysis of Eqs.
(\ref{fesr2},\ref{truevalues2}).

At this $s_0^*= 2.350\ \mathrm{GeV}^2 $, on the other hand, one finds from
Eq.~(\ref{works2}) that the OPE contribution is given by
\begin{eqnarray}\label{result1}
    a_6 + b_6 \log \frac{s_0^*}{\Lambda^2}+ \frac{b_8}{s_0^*}-
    \frac{b_{10}}{2 s_0^*{^2}}&=&\int_0^{s_0^*} dt\
    t^2 \rho(t) \\
    &=& -0.00251 \ \mathrm{GeV}^6\ ,\nn
\end{eqnarray}
whereas the exact number for the combination on the left-hand side of
Eq.~(\ref{result1}) obtained in the model is $-0.00281\ \mathrm{GeV^2}$, i.e. a
$\sim 10\%$ error.\footnote{We have repeated the analysis including the true values
for the coefficients $b_6$ and $b_8$ in the fit to Eqs. (\ref{works0},\ref{works1}).
In this case the error becomes 14\%.} The same analysis can be repeated for the
combinations appearing in the rest of Eqs.~(\ref{works3}, \ref{works5}). For
instance, we obtain
\begin{eqnarray}\label{result2}
  && \!\!\!\!\!\!\!\!\!\!\!\!\!\!\!\!\!\!\!\! a_8 +
   b_8 \log \frac{s_0^*}{\Lambda^2}- b_6\,s_0^*+ \frac{b_{10}}{s_0^*}\\
    && \qquad \qquad\qquad \qquad  =- \int_0^{s_0^*}\!\!\!\! dt\ t^3 \rho(t)  =
     0.00329\ \mathrm{GeV}^8  \quad (s_0^*= 2.307\ \mathrm{GeV}^2)\ ,\nn\\
&&\!\!\!\!\!\!\!\!\!\!\!\!\!\!\!\!\!\!\!\!\!\!\!\mathrm{and}\nn \\
 && \!\!\!\!\!\!\! \!\!\!\!\!\!\!\!\!\!\!\!\!\!\!\! a_{16} +
 b_{16} \log \frac{s_0^*}{\Lambda^2}- \frac{b_6 {s_0^*}^5}{5}+ \frac{b_{8}{s_0^*}^4}{4}
 - \frac{b_{10}{s_0^*}^3}{3}
 + \frac{b_{12}{s_0^*}^2}{2}- b_{14}s_0^* + \frac{b_{18}}{s_0^*}\label{result3}\\
 && \qquad \qquad \qquad \qquad = - \int_0^{s_0^*}\!\!\!\!\! dt\ t^7 \rho(t) = 0.0179 \
 \mathrm{GeV}^{16} \quad (s_0^*= 2.130\ \mathrm{GeV^2})\ ,\nn
\end{eqnarray}
to be compared to the exact values  on the left-hand sides of Eqs.~(\ref{result2})
and (\ref{result3}) which are $0.00344\ \mathrm{GeV^8}$ and $0.0161\
\mathrm{GeV^{16}}$, representing $\sim 4\%$ and $\sim 11\%$ errors, respectively.
The gain in precision with respect to previous methods is clear.

\section{Conclusion}

We have developed a model of duality violations that shares many properties with
QCD, including the fact that resonances in the real world have a non-vanishing
width.   Of course, an analysis of duality violations in a model cannot replace a
similar analysis in QCD.  For example, the precision of the determination  of
combinations of OPE coefficients in Eqs.~(\ref{result1}-\ref{result3}) is not
necessarily an indication of what can be achieved in QCD.  However, we do believe
that the lessons from our model analysis are also relevant for   QCD. First, the
properties of the model that lead to duality violations are also present in QCD
\cite{Shifman}, notably the fact that the OPE in both is (at best) asymptotic.
Second, because of the semi-realistic properties of the model, our analysis gives a
good indication of the nature of systematic errors which result from making the
various {\it ans\"atze} which have been proposed in the literature in order to deal
with duality violations.  We think that it should prove useful to attempt an
analysis of the duality violations represented by the ${\cal D}^{[n]}$ (cf.
Eqs.~(\ref{works0}-\ref{works5})), using an {\it ansatz} similar to that of
Eq.~(\ref{fit}), also in the QCD case.  In addition, we would like to call the
reader's attention to the additional contributions to the ``OPE side" of duality
relations, represented by the $b_{2n}$ in Eqs.~(\ref{works0}-\ref{works5}).  While
some of the $b_{2n}$ may be small in the case of QCD, there are contributions
proportional to positive powers of $s_0$, which enhance the effects of the $b_{2n}$
coefficients. To the best of our knowledge, this issue has not been taken into
account in any of the previous determinations in the literature of higher-dimension
OPE coefficients from duality. In a realistic analysis of the data, the procedure
probably should be refined  to determine all the $a$ and $b$ coefficients of the
OPE, as well as the parameters $\kappa, \gamma, \alpha$ and $\beta$ of
Eq.~(\ref{fit}) parameterizing the duality violations, by means of a simultaneous
fit to all Eqs.~(\ref{works0}-\ref{works5}), in a manner similar to that of
Ref.~\cite{Cirigliano}.

Finally, in real life one will also have to deal with experimental error bars, of
course. For example, the experimental error bars in Fig.~(\ref{fig3}) are large in
the large-$s_0$ region, and this will further complicate a precision determination
of OPE coefficients. However, we hope to have made it clear that much can be learned
from considering models such as the one analyzed in this paper, and that some of the
lessons can be applied to the real-world case of QCD.  At the very least, an
analysis like the one presented here offers a way to estimate systematic effects.

 \section*{Acknowledgements}

We would like to thank Hans Bijnens, Vincenzo Cirigliano, Eduardo de Rafael, John
Donoghue, Matthias Jamin, Jose Ignacio Latorre, Ximo Prades and Joan Rojo for
discussions. MG thanks the IFAE at the Universitat Aut\`onoma de Barcelona, and SP
thanks the Dept. of Physics at San Francisco State University for hospitality. All
three of us thank the Benasque Center for Science for hospitality. OC and SP are
supported by CICYT-FEDER-FPA2002-00748, 2001-SGR00188 and by TMR, EC-Contract No.
HPRN-CT-2002-00311 (EURIDICE).   MG is partially supported by the US Dept. of
Energy.

\section*{Appendix}

The OPE of our model stems from the asymptotic expansion of the $\psi(z)$ function
in Eq.~(\ref{asympt}). Using the identity for Bernoulli numbers
\begin{equation}\label{id}
    B_{2n}=2\ (-1)^{n+1}\  \frac{(2n)!}{(2\pi)^{2n}}\ \zeta(2n)\quad , \quad
    n=1,2,3,...
\end{equation}
where $\zeta(2n)$ is Riemann's zeta function, one can estimate the minimal error,
$\delta$, for this expansion as
\begin{equation}\label{error}
    \delta \sim \frac{B_{2n^*}}{2n^* z^{2n^*}}\
\end{equation}
where $n^*$ is the optimal value in the sum (\ref{asympt}). This $n^*$, in turn, can
be estimated as the value at which two consecutive terms in the sum (\ref{asympt})
are approximately equal. This results in  $n^*\sim \pi z$.

Consequently, after using Stirling's expression for the factorial, one obtains for
the minimum error in the asymptotic expansion (\ref{asympt})
\begin{equation}\label{finalerror}
    \delta \sim e^{-2\pi z}\ ,
\end{equation}
where, here, $z \sim \left(\frac{|q^2|}{\Lambda^2}\right)^{\zeta}$ at large $|q^2|$,
and $\zeta=1-\frac{a}{\pi N_c}$ for finite $N_c$. The estimate (\ref{finalerror}) is
also valid for the function $\Pi_{VV}-\Pi_{AA}$, because there is no cancelation in the difference
after inclusion of the right pre-factors. The result
(\ref{finalerror}) is what appears in Eqs.~(\ref{DVinf}) and (\ref{DV}) in the main
text.

We use the opportunity to also give some details of the derivation of Eq.~(\ref{c}).
{}From its definition (\ref{const}), ${\cal D}^{[n]}_{const.}(s_0)$ can be written
as
\begin{equation}
\mathcal{D}^{[n]}_{const.}(s_0)=\mathcal{C}^{[n]} + \frac{1}{2\pi i}\left\{\int_{-i\infty}^{-is_0}
   +\int_{is_0}^{i\infty}\right\}
   dq^2\ q^{2n}\ \Delta(q^2)\ .
\end{equation}
The two integrals can be estimated by using the large-$q^2$ behavior of $\Delta(q^2)$,
if $s_0$ is large.  Referring back to Eq.~(\ref{deltaa}), we note that
\begin{equation}
\label{asymp}
\cot\left[\pi\left(\frac{-q^2}{\Lambda^2}\right)^\zeta+\pi\frac{m^2}{\Lambda^2}\right]
=2e^{-2y}\ \sin(2x)\pm i\left(1+2e^{-2y}\ \cos(2x)\right)+{\cal O}\left(e^{-4y}\right)\ ,
\end{equation}
where the plus (minus) sign should be used on the positive (negative) imaginary axis, and
where we defined
\begin{eqnarray}
x&=&\pi\left|\frac{q^2}{\Lambda^2}\right|^\zeta
\cos(\zeta\pi/2)+\pi\frac{m^2}{\Lambda^2}\ ,\\
y&=&\pi\left|\frac{q^2}{\Lambda^2}\right|^\zeta\sin(\zeta\pi/2)\ .\nonumber
\end{eqnarray}
It is then straightforward to estimate each integral by using the bounds $|\sin(2x)|\le 1$,
$|\cos(2x)|\le 1$,
and taking $\zeta\to 1$.  The  unsuppressed $\pm i$ in Eq.~(\ref{asymp}) cancels
between the vector and axial channels, and we obtain the result given in Eq.~(\ref{c}).

\end{document}